\documentclass[aps,pra,reprint,superscriptaddress,amsfonts,amsmath,amssymb,longbibliography]{revtex4-2}
\usepackage{xcolor}

\usepackage{hyperref}
\usepackage{graphicx}
\usepackage{bm}
\usepackage{newtxtext}
\usepackage[varg]{newtxmath}
\usepackage[normalem]{ulem}
\usepackage{lipsum}
\usepackage{slashed}
\usepackage{mathrsfs}
\usepackage{blkarray}
\usepackage{mleftright}
\mleftright

\hypersetup{colorlinks=true,linkcolor=blue,citecolor=blue,urlcolor=blue}

\begin{document}
\title{Electron dynamics induced by quantum cat-state light}
\author{Shohei Imai}
\affiliation{Department of Physics, University of Tokyo, Hongo, Tokyo 113-0033, Japan}
\author{Atsushi Ono}
\affiliation{Department of Physics, Graduate School of Science, Tohoku University, Sendai 980-8578, Japan}
\author{Naoto Tsuji}
\affiliation{Department of Physics, University of Tokyo, Hongo, Tokyo 113-0033, Japan}
\affiliation{RIKEN Center for Emergent Matter Science (CEMS), Wako 351-0198, Japan}
\date{\today}

\begin{abstract}
We present an effective theory for describing electron dynamics driven by an optical external field in a Schr\"{o}dinger's cat state.
We show that the reduced electron density matrix evolves as an average over trajectories $\{\rho_\alpha\}$ weighted by the Sudarshan--Glauber $P$ distribution $P(\alpha)$ in the weak light--matter coupling regime.
Each trajectory obeys an equation of motion, $\mathrm{i} \partial_t\rho_\alpha=\mathcal{H}_{\alpha} \rho_\alpha-\rho_\alpha\mathcal{H}_{\alpha}$, where an effective Hamiltonian $\mathcal{H}_{\alpha}$ becomes non-Hermitian due to quantum interference of light.
The optical quantum interference is transferred to electrons through the asymmetric action between the ket and bra state vectors in $\rho_{\alpha}$.
This non-Hermitian dynamics differs from the conventional one observed in open quantum systems, described by $\mathrm{i} \partial_t\rho=\mathcal{H}\rho-\rho \mathcal{H}^\dagger$, which has complex conjugation in the second term.
We confirm that the reduced, trajectory-resolved effective theory agrees with full electron-photon simulations for the few-electron Dicke model, thereby validating the interferential non-Hermitian description in the weak-coupling regime.
\end{abstract}
\maketitle

\textit{Introduction}---%
Quantum superposition of macroscopically distinct states, exemplified by Schr\"{o}dinger's cat state, has long symbolized the mysterious properties of quantum mechanics~\cite{Schrodinger1935, Wineland2013, Frowis2018}.
Such quantum superpositions are essential for enhancing computational speed in quantum computing~\cite{Shimizu2013, Aberg2006, Stahlke2014, Theurer2017, Chitambar2019}, and will play a fundamental role in controlling large-scale quantum systems.
Despite their conceptual importance, generating large-scale quantum cat states has been experimentally challenging.
Recently, several breakthroughs have emerged in attosecond science~\cite{Lewenstein2021, Rivera-Dean2021, Stammer2023, Cruz-rodriguez2024, Lamprou_2025, Gonoskov2024}, synchrotron radiation technique~\cite{Dahan2023}, and photonic quantum information technology~\cite{Konno2024a}, which pave the way for experimentally realizing large-scale optical cat states.
These advances are opening new avenues for measuring, controlling, and engineering quantum materials through quantum light irradiation~\cite{Stammer2024b, Kira2011, Lamprou2025, Dorfman2016, Mukamel2017, Lamprou2020, Lysne2021a, Fedotov2023, Sennary2024}.

From a theoretical point of view, it has been challenging to analyze electron dynamics under such large-scale quantum light irradiation.
First, it is computationally expensive to simulate entire light--matter coupled systems, including quantized photons.
Second, an effective theory for open quantum systems such as the Gorini--Kossakowski--Sudarshan--Lindblad formalism~\cite{Gorini1975, Lindblad1976} often relies on the Markov approximation, which disrupts optical quantum interference.
This formalism has been used, for instance, in the analysis of atomic gases driven by broadband squeezed vacuum states~\cite{Agarwal1990}.
Recently, there have been proposed non-Markovian effective theories for quantum-light-driven systems~\cite{Gorlach2023, Tzur2022a, Khalaf2023, Rasputnyi2024a, EvenTzur2024, 4x11-phs2}, where the reduced density matrix of electrons driven by quantum light is described by electron state vectors driven by classical light.
Here, we pose a question: Are there any characteristic electron dynamics caused by optical quantum interference that cannot be reproduced with such classical-light reduction?

In this study, we propose an alternative effective theory for describing quantum-light-driven electron dynamics in the weak light--matter coupling regime that does not rely on the Markov approximation and classical-light reduction, using the path integral and Born approximation.
This is achieved by representing the quantum state of light using the Sudarshan--Glauber $P$ representation~\cite{Sudarshan1963, Glauber1963a}.
We show that optical quantum interference introduces a contribution to the reduced electron density matrix that obeys unconventional non-Hermitian dynamics with a complex-valued electromagnetic field.
Through the analysis of a two-qubit electron system coupled to a photon field, we verify the validity of our effective theory, and observe that quantum interference of cat-state light is transferred to that of electrons via the effective non-Hermitian dynamics.
We also compare our theory with the previous formalism~\cite{Gorlach2023, Tzur2022a, Khalaf2023, Rasputnyi2024a, EvenTzur2024, 4x11-phs2}, highlighting the role of the non-Hermitian dynamics in generating quantum superposition and quantum interference in electron systems.

\textit{Formalism}---%
We first consider the following general Hamiltonian:
\begin{equation}
\hat{\mathcal{H}} = \hat{\mathcal{H}}_{\mathrm{e}}[\gamma\hat{a}] + \hat{\mathcal{H}}_{\mathrm{p}}[\hat{a}]. \label{eq:hamiltonian}
\end{equation}
Here, $\hat{a}$ represents the bosonic annihilation operator of the photon field.
The first term $\hat{\mathcal{H}}_{\mathrm{e}}$ denotes the Hamiltonian for matter, which is coupled to the photon $\hat{a}$ with a coupling constant $\gamma$.
Here, we focus on an electron system (the following derivation can be applied to other charged systems as well).
The second term $\hat{\mathcal{H}}_{\mathrm{p}}$ represents the photon Hamiltonian.
As will be shown below, any form of the Hamiltonian can be considered, provided that the equation of motion for the free photon field is analytically solvable.
For simplicity, we assume a harmonic oscillator with frequency $\omega$, namely $\hat{\mathcal{H}}_{\mathrm{p}} = \omega \hat{a}^{\dagger} \hat{a}$.
We further assume that the bosonic operators $\hat{a}$ in both $\hat{\mathcal{H}}_{\mathrm{e}}$ and $\hat{\mathcal{H}}_{\mathrm{p}}$ are in the normal order.
The Dirac constant $\hbar$ and the electric charge $q$ are set to unity.

Light irradiation is introduced by setting the initial state of the photon field to a nonequilibrium state.
Here, we express an arbitrary initial density matrix of photons as $\hat{\rho}_{\mathrm{p}}(0)$ (with the initial time set to $0$), using the Sudarshan--Glauber $P$ representation~\cite{Sudarshan1963, Glauber1963a}.
In this representation, any density matrix of a boson field can be expressed in terms of a real-valued quasiprobability distribution function $P(\alpha)$ as
\begin{align}
&\hat{\rho}_{\mathrm{p}}(0) = \int \mathrm{d}^2 \alpha P(\alpha) |\alpha\rangle\langle \alpha|, \label{eq:Sudarshan--Glauber P representation} \\
&|\alpha\rangle = \hat{D}(\alpha)|0\rangle,\ \ \ \ \hat{D}(\alpha) = \mathrm{e}^{\alpha \hat{a}^{\dagger} - \overline{\alpha} \hat{a}}, \label{eq:coherent state}
\end{align}
where $\hat{\rho}_{\mathrm{p}}(0)$ is represented in the diagonal form in a coherent state $|\alpha\rangle$, which is a classical state corresponding to a laser.
Here, $\overline{\alpha}$ denotes the complex conjugate of $\alpha$.
The advantage of this representation is that one can directly employ the path integral and Born approximation, thanks to the integral form of the classical density matrix $|\alpha\rangle\langle\alpha|$, as discussed below.

The density matrix of electrons at time $t$ can be expressed with a path integral representation over the photon modes~\cite{Feynman:100771} as follows:
\begin{align}
\hat{\rho}_{\mathrm{e}}(t) =& \int \mathrm{d}^2\alpha P(\alpha) 
\int_{\alpha_+(0) = \alpha_-(0) = \alpha}^{\alpha_+(t) = \alpha_-(t) }
\mathcal{D}[\alpha_+,\alpha_-] \nonumber \\
&\ \ \ \ \ \times \hat{\mathrm{T}}_+\, \mathrm{e}^{\mathrm{i} \hat{S}_{+}(t)}  \, 
\hat{\rho}_{\mathrm{e}}(0) \,
\hat{\mathrm{T}}_-\, \mathrm{e}^{-\mathrm{i} \hat{S}_{-}(t)}, \label{eq:path integral}
\end{align}
where $\hat{S}_{\lambda=\pm}(t)$ is the action defined by
\begin{equation}
\hat{S}_{\lambda}(t) = 
\int_{0}^{t}\mathrm{d}\tau 
\left\{\overline{\alpha_{\lambda}}\mathrm{i}\partial_{\tau} \alpha_{\lambda}
- \mathcal{H}_{\mathrm{p}}[\alpha_{\lambda}] 
- \hat{\mathcal{H}}_{\mathrm{e}}[\gamma \alpha_{\lambda}] \right\}. \label{eq:action}
\end{equation}
Here, we use Eq.~\eqref{eq:Sudarshan--Glauber P representation} and trace out the photon degrees of freedom.
$\hat{\mathrm{T}}_+$ ($\hat{\mathrm{T}}_-$) denotes the (anti)time-ordering, and $\lambda\, (= +,-)$ indicates the Keldysh paths, where $\alpha_{\lambda}(\tau)$ represents the classical trajectory of the photon field.
In the initial state, we assume that electrons and photons are separable, i.e., $\hat{\rho}_{\mathrm{e}}(0) \otimes \hat{\rho}_{\mathrm{p}}(0)$.
The interaction term $\hat{\mathcal{H}}_{\mathrm{e}}$ within the action~\eqref{eq:action} includes not only the effect of nonequilibrium photons driving the electrons, but also the back-action from the electrons onto the photon field, generally leading to intertwined dynamics.

When the electron--photon coupling constant $\gamma$ is sufficiently small, we can use an iterative approximation (i.e., the Born approximation) that separates the electron dynamics from the photon dynamics.
Setting $\gamma = 0$ in the action~\eqref{eq:action} yields the equation of motion for the free photon field as
\begin{equation}
\mathrm{i} \partial_{\tau} \alpha(\tau) = \frac{\updelta \mathcal{H}_{\mathrm{p}} }{\updelta \overline{\alpha}}[\alpha(\tau)], \label{eq:0th-order photon EoM}
\end{equation}
where the index for the Keldysh paths is omitted.
The solution of this equation is denoted as $\alpha_0(\tau)$.
For example, in the case of the harmonic oscillator $\hat{\mathcal{H}}_{\mathrm{p}} = \omega \hat{a}^{\dagger} \hat{a}$, we obtain the solution for the initial condition $\alpha_0(0) = \alpha$ as
\begin{equation}
\alpha_0(\tau) = \alpha \mathrm{e}^{-\mathrm{i}\omega \tau}. \label{eq:free photon field}
\end{equation}

\begin{figure}[bt]
\centering
\includegraphics[width=1.0\columnwidth]{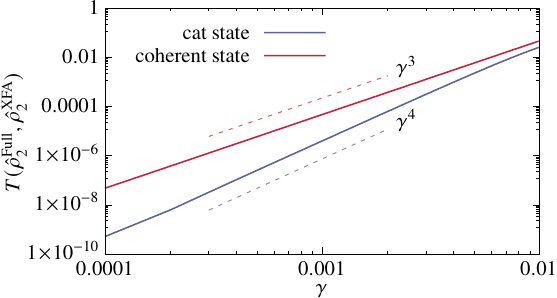}
\caption{Trace distance between the density matrices of the full simulation and the external-field approximation (XFA) simulation at time $t=100$ and amplitude $\alpha_0 = 0.8$.}
\label{fig:gamma}
\end{figure}

We proceed to the dynamics of the electron system driven by the free photon field $\alpha_0(\tau)$. 
Within the Born approximation, we only consider the free trajectory $\alpha_0(\tau)$ among all the classical field trajectories to be summed over in the path integral~\eqref{eq:path integral}.
Then, the electron dynamics can be described by the following density matrix and equations of motion:
\begin{align}
&\hat{\rho}_{\mathrm{e}}(t) \approx \int \mathrm{d}^2 \alpha P(\alpha) \hat{\rho}_{\mathrm{e},\alpha}(t), \label{eq:P weighted density matrix} \\
&\mathrm{i}\partial_{t} \hat{\rho}_{\mathrm{e},\alpha}(t) = \left[\hat{\mathcal{H}}_{\mathrm{e}}[\gamma \alpha_0(t)],\hat{\rho}_{\mathrm{e},\alpha}(t) \right]. \label{eq:EoM}
\end{align}
Equations~\eqref{eq:P weighted density matrix} and~\eqref{eq:EoM} are central formulas of this paper, indicating that the electron density matrix $\hat{\rho}_{\mathrm{e}}(t)$ is described by an average of $\hat{\rho}_{\mathrm{e},\alpha}(t)$ (trajectories), each of which is weighted by the photon's quasiprobability distribution $P(\alpha)$ and is governed by the von Neumann equation with a Hamiltonian where the bosonic operator $\hat{a}$ is replaced with the scalar field $\alpha_0(t)$.
At this stage, we do not take account of the electron back-action on photons, and treat the photon field as an external field---a treatment hereafter referred to as the external-field approximation.
The Born approximation is formulated as an expansion in the electron--photon coupling constant $\gamma$, where the back-action on the photon field $\alpha_0(\tau)$ enters through a term proportional to $\gamma \mathcal{J}$, with $\mathcal{J}$ denoting the electric current [see Fig.~\ref{fig:gamma} and Eq.~\eqref{eq:1st-order photon EoM} for details].
Meanwhile, the action on the electron system always enters in the form $\gamma \alpha$, suggesting that the external-field approximation remains reliable as long as this quantity is small.
The time evolution of the density matrix $\hat{\rho}_{\mathrm{e}}(t)$ is non-Markovian, in the sense that it cannot be determined by a closed form of a differential equation involving only $\hat{\rho}_{\mathrm{e}}(t)$.

Let us apply the present formalism to the even Schr\"{o}dinger cat-state light defined as
\begin{equation}
|\mathrm{cat}\rangle = \frac{1}{\sqrt{\mathcal{N}_{\alpha_0}}} \left( |\alpha_0\rangle + |{-}\alpha_0\rangle \right), \label{eq:cat state}
\end{equation}
where  $\mathcal{N}_{\alpha_0}=2[1+\langle-\alpha_0|\alpha_0\rangle] = 2[1+\exp(-2|\alpha_0|^2)]$ is the normalization constant.
The corresponding Sudarshan--Glauber $P$ function is given by
\begin{align}
&P_{\mathrm{cat}}(\alpha) = \frac{1}{\mathcal{N}_{\alpha_0}} \Bigl\{ \delta^2(\alpha - \alpha_0) + \delta^2(\alpha + \alpha_0) \nonumber \\
&+ \langle -\alpha_0|\alpha_0 \rangle \left[ \tilde{\delta}(\alpha{-}\alpha_0)\tilde{\delta}(\overline{\alpha}{+}\overline{\alpha_0}) + \tilde{\delta}(\alpha{+}\alpha_0)\tilde{\delta}(\overline{\alpha}{-}\overline{\alpha_0}) \right] \Bigr\}, \label{eq:cat state:P}
\end{align}
where we use the generalized delta function~\cite{Brewster2018} defined as $\int \mathrm{d}^2\alpha f(\alpha,\overline{\alpha})\, \tilde{\delta}(\alpha - \alpha_1) \tilde{\delta}(\overline{\alpha}-\overline{\alpha_2}) = f(\alpha_1, \overline{\alpha_2})$ for an arbitrary complex function $f(\alpha,\overline{\alpha})$.
The first and second terms in Eq.~\eqref{eq:cat state:P} correspond to classical states, specifically $|\alpha_0\rangle \langle\alpha_0|$ and $|{-}\alpha_0\rangle \langle-\alpha_0|$, indicating a mapping of the photon operators $(\hat{a},\hat{a}^{\dagger})$ to $(\alpha_0,\overline{\alpha_0})$ and $(-\alpha_0,-\overline{\alpha_0})$, respectively.
Meanwhile, the third and fourth terms in Eq.~\eqref{eq:cat state:P} represent quantum interference terms, namely $|\alpha_0\rangle \langle-\alpha_0|$ and $|{-}\alpha_0\rangle \langle\alpha_0|$, replacing $(\hat{a},\hat{a}^{\dagger})$ with $(\alpha_0,-\overline{\alpha_0})$ and $(-\alpha_0,\overline{\alpha_0})$, respectively.
Consequently, $\alpha$ and $\overline{\alpha}$ are no longer complex conjugates of each other, causing physical quantities such as the electric field and vector potential to become complex numbers in the interference terms.
Of particular interest is the contribution driven by the interference terms, where the electron Hamiltonian in Eq.~\eqref{eq:EoM} becomes non-Hermitian due to the breakdown of the Hermitian conjugacy between $\alpha$ and $\overline{\alpha}$.

This non-Hermiticity differs from what is typically observed in Markovian open systems, as the assumed equation of motion takes a different form.
Let us write the system's density matrix as $\hat{\rho}_{\mathrm{s}}$ and the non-Hermitian Hamiltonian as $\hat{\mathcal{H}}_{\mathrm{nh}}$. 
In the Gorini--Kossakowski--Sudarshan--Lindblad formalism~\cite{Gorini1975, Lindblad1976}, the equation of motion is given by $\mathrm{i} \partial_t \hat{\rho}_{\mathrm{s}} = \hat{\mathcal{H}}_{\mathrm{nh}}\hat{\rho}_{\mathrm{s}} - \hat{\rho}_{\mathrm{s}} \hat{\mathcal{H}}_{\mathrm{nh}}^{\dagger}$, where quantum jump terms are neglected.
This describes dynamics that do not conserve the trace of $\hat{\rho}_{\mathrm{s}}$~\cite{Hatano1996, Bender1998, Brody2014, Szankowski2023, Felski2024}, referred to here as \textit{dissipative} non-Hermitian dynamics.
In contrast, when we consider the contribution driven by the quantum interference of light, our equation of motion [Eq.~\eqref{eq:EoM}] takes the form of $\mathrm{i} \partial_t \hat{\rho}_{\mathrm{s}} = \hat{\mathcal{H}}_{\mathrm{nh}}\hat{\rho}_{\mathrm{s}} - \hat{\rho}_{\mathrm{s}} \hat{\mathcal{H}}_{\mathrm{nh}}$ with $\hat{\rho}_{\mathrm{s}}$ being replaced by $\hat{\rho}_{\mathrm{e},\alpha}$.
An important difference from the dissipative non-Hermitian dynamics is that the actions on the ket and bra spaces are not Hermitian conjugates of each other, resulting in the operator $\hat{\rho}_{\mathrm{e},\alpha}$ itself becoming non-Hermitian (while its trace remains conserved).
This indicates that $\hat{\rho}_{\mathrm{e},\alpha}$ evolves into a quantum interference term, referred to here as \textit{interferential} non-Hermitian dynamics.
The resulting operator $\hat{\rho}_{\mathrm{e}}$ becomes Hermitian after taking the average in Eq.~\eqref{eq:P weighted density matrix}.
Note that the full electron-photon Hamiltonian remains Hermitian; the non-Hermiticity arises only when considering the trajectory-resolved reduced description of the cat state's interference sectors.

This reduced description has operational significance because the interference-sector contributions are encoded in measurable photon observables.
In a simplified form, the electron density matrix is expressed by the total density matrix $\hat{\rho}_{\mathrm{tot}}$ as $\hat{\rho}_{\mathrm{e}}=\mathrm{Tr}_{\mathrm{p}}[\hat{\rho}_{\mathrm{tot}}]=\hat{\rho}_{++}+\hat{\rho}_{--}+\hat{\rho}_{+-}+\hat{\rho}_{-+}$, where each term corresponds to those in Eq.~\eqref{eq:cat state:P}, respectively, up to normalization.
Using the expectation values of the photon field operators $\hat{a}$, $\hat{a}^{\dagger}$, and $\hat{a}^{\dagger}\hat{a}$, we have $\mathrm{Tr}_{\mathrm{p}}[\hat{\rho}_{\mathrm{tot}}(t)\hat{a}]/\alpha_0(t)=\hat{\rho}_{++} - \hat{\rho}_{--} + \hat{\rho}_{+-} - \hat{\rho}_{-+}$, $\mathrm{Tr}_{\mathrm{p}}[\hat{\rho}_{\mathrm{tot}}(t)\hat{a}^{\dagger}]/\overline{\alpha_0(t)} = \hat{\rho}_{++} - \hat{\rho}_{--} - \hat{\rho}_{+-} + \hat{\rho}_{-+}$, and $\mathrm{Tr}_{\mathrm{p}}[\hat{\rho}_{\mathrm{tot}}(t)\hat{a}^{\dagger}\hat{a}]/|\alpha_0(t)|^2 = \hat{\rho}_{++} + \hat{\rho}_{--} - \hat{\rho}_{+-} - \hat{\rho}_{-+}$.
These expressions allow one to explicitly extract the contributions of the quantum interference terms ($\hat{\rho}_{+-}$ and $\hat{\rho}_{-+}$) from physical observables.

\textit{Two-electron system}---%
To validate our effective theory [Eqs.~\eqref{eq:P weighted density matrix} and~\eqref{eq:EoM}], we analyze the widely used two-qubit--one-boson coupled model, called the two-qubit Dicke or Rabi model~\cite{Dicke1954, Rabi1937, Hanamura1993, Ficek2002, Li2004a, PRAKASH2008, Chilingaryan2013, Combescot2015, Mohamed2019a, Mohamed2022, Eshun2022, Movahedi2023}, and compare the results with full system simulations.
The Hamiltonian is given by
\begin{equation}
\hat{\mathcal{H}}_{\mathrm{e}}[\gamma \hat{a}] = \sum_{j=1,2} \left( \varDelta \hat{S}^{z}_{j} - \hat{E} \cdot \mu \hat{S}^x_j \right), \ \hat{E}=\mathrm{i}\gamma\omega(\hat{a}-\hat{a}^{\dagger}),\label{eq:ham:2spin--1boson}
\end{equation}
where $\hat{\bm{S}}_j$ represents the pseudo-spin-$1/2$ operators that describe the effective electron states modeled as a two-level system, such as an exciton, and the subscript $j=1,2$ identifies each exciton.
Here, $\varDelta$ is the excitonic binding energy, and $\mu$ is the electric dipole moment.
The photon Hamiltonian is $\hat{\mathcal{H}}_{\mathrm{p}} = \omega \hat{a}^{\dagger} \hat{a}$, and $\hat{E}$ is its electric field.
Taking $\varDelta = 1$ as the energy unit, we assume the resonance condition $\omega = \varDelta$, absorb the dipole moment into the coupling constant by setting $\mu = 1$, and set the dimensionless coupling strength to a sufficiently small value, $\gamma = 10^{-3}$.
In this setup, our effective theory is at most valid up to a typical light--matter interacting timescale $t_{\mathrm{c}} \sim 1/\mu \omega \gamma=1000$.

We solve the model numerically as follows.
First, we calculate the time evolution of the entire system’s density matrix $\hat{\rho}^{\mathrm{Full}}(t)$ by solving the von Neumann equation $\mathrm{i} \partial_t \hat{\rho}^{\mathrm{Full}}(t) = [\hat{\mathcal{H}},\hat{\rho}^{\mathrm{Full}}(t)]$ using the fourth-order Runge--Kutta method, where $\hat{\mathcal{H}}$ in Eq.~\eqref{eq:hamiltonian} is given by Eq.~\eqref{eq:ham:2spin--1boson} and $\hat{\mathcal{H}}_{\mathrm{p}} = \omega \hat{a}^{\dagger} \hat{a}$.
Here, the initial state of electrons is set to the ground state of $\hat{\mathcal{H}}_{\mathrm{e}}[0]$.
We truncate the photon Hilbert space such that the maximum photon number is $100$, for which we confirm that the numerical results are sufficiently converged.
For the effective theory under the external-field approximation, each von Neumann equation~\eqref{eq:EoM} is solved by the fourth-order Runge--Kutta method.
The time step is set to $\updelta t = 0.01$.

\begin{figure}[bt]
\centering
\includegraphics[width=\columnwidth]{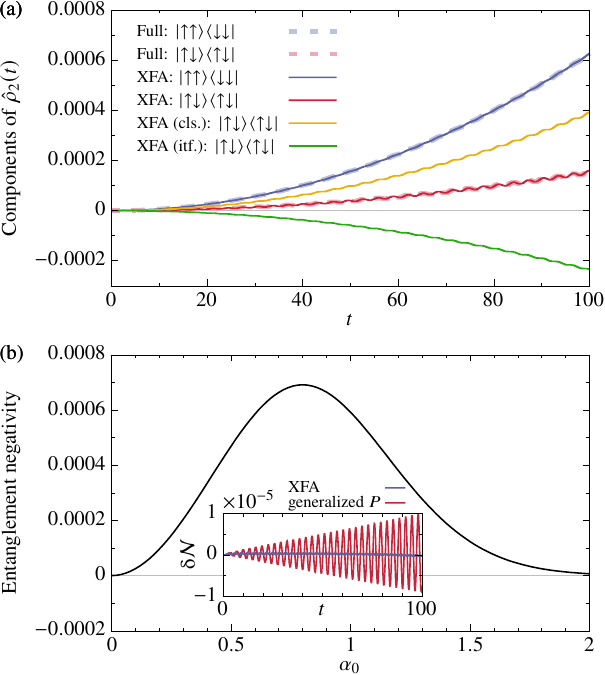}
\caption{(a)~Dynamics of the two-body density matrix $\hat{\rho}_2(t)$ for $\alpha_0 = 0.5$.
Cls.\ and itf.\ stand for classical and interferential components, respectively.
XFA refers to the external-field approximation.
(b)~Entanglement negativity between electrons.
The inset compares the time evolution of the deviation in entanglement negativity from the full-system simulation for our effective theory (blue) and the generalized $P$-representation theory (red).
}
\label{fig:cat}
\end{figure}
Figure~\ref{fig:cat}(a) shows the dynamics of non-zero components in the two-body density matrix $\hat{\rho}_2(t)$ for two electrons.
With $|{\uparrow}\rangle$ and $|{\downarrow}\rangle$ being eigenstates of each electron’s $\hat{S}^z_j$, the absolute value of the corner component $|{\uparrow\uparrow}\rangle \langle{\downarrow\downarrow}|$ in $\hat{\rho}_2(t)$ is plotted in blue, while the component of $|{\uparrow\downarrow}\rangle \langle{\uparrow\downarrow}|$ corresponding to the single-electron excitation is plotted in red.
The dashed curves are the results from the full-system simulation, while the solid curves are those from the effective theory.
There is a good agreement between them, indicating that the external-field approximation and our effective theory work well.

Let us consider the physical interpretation of these results.
The expectation value of the electric polarization $\hat{S}_{1,2}^+$ is zero because the electric field and vector potential vanish in the even cat state~\eqref{eq:cat state}.
In contrast, the expectation value of $\hat{S}_1^+ \hat{S}_2^+ = |{\uparrow\uparrow}\rangle \langle{\downarrow\downarrow}|$, which corresponds to biexciton polarization, is finite.
This indicates the generation of a coherent biexciton.
The small oscillation observed in every curve is due to the contributions of the counter-rotating terms.
The effect of quantum interference appears rather in the single-electron excitation, exemplified by the component of $|{\uparrow\downarrow}\rangle \langle{\uparrow\downarrow}|$.
According to Eq.~\eqref{eq:cat state:P}, the electron dynamics driven by the cat-state light is decomposed into four contributions: the first two correspond to classical field driving, while the third and fourth are driven by the interference terms.
These two contributions are plotted in Fig.~\ref{fig:cat}(a) as the yellow and green curves, respectively (with their sum matching the red curve).
Notably, while the diagonal components of the density matrix are always positive due to the probability interpretation, the interferential non-Hermitian driving results in a negative contribution (the green curve) to the single-electron excitation, thereby suppressing the total single-electron excitation.

Furthermore, the cat-state light generates entanglement between the two electrons.
In our effective electron theory, this entanglement can be attributed to the interferential non-Hermitian driving.
This is because, in the present noninteracting-electron system, Hermitian classical-field driving cannot generate entanglement, consistent with the local operations and classical communication (LOCC) criterion.
To analyze entanglement in a density matrix, we examine the Peres--Horodecki criterion~\cite{Peres1996, Horodecki1996, Vidal2002, Horodecki2009}:
when the negativity $\mathcal{N}(\hat{\rho}_2^{\mathrm{PT}}) \equiv (\Vert \hat{\rho}_2^{\mathrm{PT}} \Vert - 1)/2$ of the partially transposed density matrix $\hat{\rho}_2^{\mathrm{PT}}$ is nonzero, it indicates the presence of electron entanglement.
Here, the partial transpose is defined as $|{\uparrow}\sigma_2\rangle \langle {\downarrow} \sigma_2'| \rightleftharpoons |{\downarrow} \sigma_2\rangle \langle {\uparrow} \sigma_2'|$ ($\sigma_2,\sigma_2' = {\uparrow},{\downarrow}$), and $\Vert \hat{A} \Vert$ denotes the trace norm of a matrix $\hat{A}$.
Figure~\ref{fig:cat}(b) shows the entanglement negativity $\mathcal{N}(\hat{\rho}_2^{\mathrm{PT}})$ at time $t = 100$, plotted as a function of the cat-state amplitude $\alpha_0$.
The finite negativity appears for finite amplitude $\alpha_0$, indicating that the cat-state light generates entanglement between electrons.
The entanglement decreases from $\alpha_0 \approx 0.8$, which is attributed to the prefactor $\langle -\alpha_0|\alpha_0\rangle$ in the interference term~\eqref{eq:cat state:P}, indicating that the electron state approaches a classical state as the influence of the interference term diminishes for large $\alpha_0$.

To gain more insight analytically, we perform a perturbative expansion in terms of the light amplitude.
We consider electron dynamics in the case where the photon density matrix contains the component $|\alpha_1\rangle \langle\alpha_2|$, replacing $\hat{E}$ in Eq.~\eqref{eq:ham:2spin--1boson} by the complex-valued electric field $E_{\alpha_1,\alpha_2}(t)\equiv \mathrm{i}\gamma\omega [\alpha_1 \exp(-\mathrm{i}\omega t) - \overline{\alpha_2} \exp(\mathrm{i}\omega t)]$.
Since the Hamiltonian becomes independent for each electron, we can obtain the electron state vectors under the condition $\gamma |\alpha_{1,2}|t \ll 1$ as
\begin{align}
&|\psi_{\mathrm{I}}(t)\rangle \approx 
\left(
\begin{array}{c}
\mathcal{A}(\alpha_1,\alpha_2) \\
1
\end{array}
\right),\ 
\langle \psi_{\mathrm{I}}(t)| \approx 
\left( \overline{\mathcal{A}(\alpha_2,\alpha_1)}, 1\right), \label{eq:perturb:one-body state} \\
&\mathcal{A}(\alpha_1,\alpha_2) \equiv \mathrm{i}\mu \int_0^t \mathrm{d}\tau \left[ E_{\alpha_1,\alpha_2}(\tau)\, \mathrm{e}^{\mathrm{i}\varDelta \cdot \tau} \right], \label{eq:pertub:amplitude}
\end{align}
where the analysis is performed in the interaction picture to $\hat{\mathcal{H}}_{\mathrm{e}}[0]$ in Eq.~\eqref{eq:ham:2spin--1boson}, and we define $|{\uparrow}\rangle=(1,0)^{\mathrm{T}}$ and $|{\downarrow}\rangle=(0,1)^{\mathrm{T}}$ ($\mathrm{T}$ denotes transpose).
Note that $\langle \psi_{\mathrm{I}}(t)| \neq (|\psi_{\mathrm{I}}(t)\rangle)^{\dagger}$, when $\alpha_1 \neq \alpha_2$.
Then, the two-body density matrix under the cat-state light irradiation~\eqref{eq:cat state} is separated into two terms driven by the classical and interference components as $\hat{\rho}_{2,\mathrm{I}} = \hat{\rho}_{2,\mathrm{I}}^{\mathrm{cls}} + \hat{\rho}_{2,\mathrm{I}}^{\mathrm{itf}}$, each of which is given by
\begingroup
\allowdisplaybreaks
\begin{align}
\hat{\rho}_{2,\mathrm{I}}^{\mathrm{cls}} \approx&
\frac{2}{\mathcal{N}_{\alpha_0}}\, 
\begin{blockarray}{ccccc}
|{\uparrow\uparrow}\rangle & |{\uparrow\downarrow}\rangle & |{\downarrow\uparrow}\rangle & |{\downarrow\downarrow}\rangle & \\
\begin{block}{(cc|cc)c}
|a|^4 & 0 & 0 & a^2 & \, \langle{\uparrow\uparrow}| \\[2pt]
0 & |a|^2 & |a|^2 & 0 & \,\langle{\uparrow\downarrow}| \\[1pt]
\cline{1-4}\\[-10pt]
0 & |a|^2 & |a|^2 & 0 & \,\langle{\downarrow\uparrow}| \\[2pt]
\overline{a}^2 & 0 & 0 & 1 & \,\langle{\downarrow\downarrow}| \\[2pt]
\end{block}
\end{blockarray} \label{eq:two-body density matrix:cls:matrix form} \\
=& \left( |\psi_+^{\mathrm{cls}}\rangle\langle\psi_+^{\mathrm{cls}}| + |\psi_-^{\mathrm{cls}}\rangle\langle\psi_-^{\mathrm{cls}}| \right)/\mathcal{N}_{\alpha_0}, \label{eq:two-body density matrix:cls:state vector} \\
\hat{\rho}_{2,\mathrm{I}}^{\mathrm{itf}} \approx& \frac{2\langle -\alpha_0|\alpha_0\rangle}{\mathcal{N}_{\alpha_0}} 
\left( \begin{array}{cc|cc}
|b|^4 & 0 & 0 & b^2 \\
0 & -|b|^2 & -|b|^2 & 0 \\ \hline
0 & -|b|^2 & -|b|^2 & 0 \\
\overline{b}^2 & 0 & 0 & 1
\end{array} \right) \label{eq:two-body density matrix:itf:matrix form} \\
=& \langle -\alpha_0|\alpha_0\rangle
\left( |\psi_+^{\mathrm{itf}}\rangle\langle\psi_-^{\mathrm{itf}}| + |\psi_-^{\mathrm{itf}}\rangle\langle\psi_+^{\mathrm{itf}}| \right) /\mathcal{N}_{\alpha_0}, \label{eq:two-body density matrix:itf:state vectors}
\end{align}
\endgroup
respectively.
Here, we define $a \equiv \mathcal{A}(\alpha_0,\alpha_0)$ and $b \equiv \mathcal{A}(\alpha_0,-\alpha_0)$, and $|\psi_{\pm}^{\mathrm{cls}}\rangle$ and $|\psi_{\pm}^{\mathrm{itf}}\rangle$ are the state vectors of electrons driven by the real-valued electric field $\pm E_{\alpha_0,\alpha_0}$ and the imaginary-valued electric field $\pm E_{\alpha_0,-\alpha_0}$ expressed as
\begin{align}
&|\psi_{\pm}^{\mathrm{cls}}\rangle = |{\downarrow\downarrow}\rangle \pm a|{\uparrow\downarrow}\rangle \pm a|{\downarrow\uparrow}\rangle +a^2|{\uparrow\uparrow}\rangle, \label{eq:pure state:cls} \\
&|\psi_{\pm}^{\mathrm{itf}}\rangle = |{\downarrow\downarrow}\rangle \pm b|{\uparrow\downarrow}\rangle \pm b|{\downarrow\uparrow}\rangle +b^2|{\uparrow\uparrow}\rangle, \label{eq:pure state:itf}
\end{align}
respectively, where the normalization is omitted.

A naive physical interpretation of the above result is as follows.
When the amplitude $\alpha_0$ is small, the prefactor of the interference terms becomes $\langle -\alpha_0|\alpha_0\rangle = \exp(-2|\alpha_0|^2) \approx 1$.
Furthermore, due to the resonance condition, the rotating-wave approximation is also valid, yielding $\mathcal{A}(\alpha_1,\alpha_2) \approx -\mu \omega \gamma \alpha_1 t$ and $|\psi_{\pm}^{\mathrm{cls}}\rangle \approx |\psi_{\pm}^{\mathrm{itf}}\rangle$.
As a result, the electron state $\hat{\rho}_{2,\mathrm{I}}$ becomes a pure state, expressed as $|\psi_+^{\mathrm{cls}}\rangle + |\psi_-^{\mathrm{cls}}\rangle$.
This demonstrates that the cat-state light realizes a quantum superposition of electrons, each driven by the constituent classical light.
Consequently, this quantum superposed state is described as $|{\downarrow\downarrow}\rangle + a^2|{\uparrow\uparrow}\rangle$, indicating that single-electron excitation is suppressed while the electrons are entangled.

In the above, $|\psi_{\pm}^{\mathrm{itf}}\rangle$ does not play any specific role since $|\psi_{\pm}^{\mathrm{itf}}\rangle \approx |\psi_{\pm}^{\mathrm{cls}}\rangle$.
However, there are situations where $|\psi_{\pm}^{\mathrm{itf}}\rangle$ significantly deviates from $|\psi_{\pm}^{\mathrm{cls}}\rangle$ (e.g., when the rotating-wave approximation breaks down).
We confirm this by comparing our results with the effective theory based on the generalized $P$ representation~\cite{Drummond1980a} proposed in Ref.~\cite{Gorlach2023}, where $|\psi_{\pm}^{\mathrm{itf}}\rangle$ does not appear.
In the inset of Fig.~\ref{fig:cat}(b), we show the time evolution of the deviation in the entanglement negativity from the full-system simulation, denoted as $\updelta \mathcal{N}$, for our effective theory (blue) and the generalized $P$-representation theory (red).
We use the same model employed to obtain Fig.~\ref{fig:cat}.
We find that the present effective theory accurately reproduces the quantum light-driven electron dynamics, implying that the interferential non-Hermitian contribution improves the reduced description.
For more details, see the Supplemental Material~\cite{SM}.

We can also examine the response to a general cat state, $| \mathrm{cat}_{\theta} \rangle \propto | \alpha_0 \rangle + \mathrm{e}^{\mathrm{i}\theta} |{-\alpha}_0 \rangle$.
In particular, for the odd cat state ($\theta = \pi$), the electron density matrix is given by $\hat{\rho}_{2,\mathrm{I}} = \hat{\rho}_{2,\mathrm{I}}^{\mathrm{cls}} - \hat{\rho}_{2,\mathrm{I}}^{\mathrm{itf}}$.
In the limit $\alpha_0 \ll 1$, we find that $\hat{\rho}_{2,\mathrm{I}} \approx |{\downarrow\downarrow}\rangle\langle{\downarrow\downarrow}| + |\mu\omega\gamma t|^2/2 \cdot (|{\uparrow\downarrow}\rangle + |{\downarrow\uparrow}\rangle)(\langle{\uparrow\downarrow}| + \langle{\downarrow\uparrow}|)$, indicating that the electron system is in a mixed state containing an entangled single-exciton component ($|{\uparrow\downarrow}\rangle + |{\downarrow\uparrow}\rangle$).
These results highlight how the quantum interference phase $\theta$ in the cat state directly influences the nature of the generated electron entanglement.
Further details are provided in the Supplemental Material~\cite{SM}.

Based on the perturbative analysis, the entanglement negativity is expressed as $|\mu\gamma\omega\alpha_0 t|^2 \langle -\alpha_0 | \alpha_0 \rangle / \mathcal{N}_{\alpha_0}$ and grows proportionally to $\gamma^2$.
However, as one increases $\gamma$, the time scale in which the effective theory is valid becomes shorter.
Determining how large entanglement can be generated, and whether our effective theory can describe it is an important open question.

\textit{Electron--photon coupling dependence}---%
We discuss the validity of our effective theory with respect to the electron--photon coupling constant $\gamma$, which serves as a control parameter in the external-field approximation.
First, we numerically calculate the deviation between $\hat{\rho}_2^{\mathrm{Full}}$ obtained from the full-system simulation and $\hat{\rho}_2^{\mathrm{XFA}}$ obtained under the external-field approximation.
In these calculations, we apply the rotating wave approximation for simplicity.
Figure~\ref{fig:gamma} shows the trace distance $T(\hat{\rho}_2^{\mathrm{Full}}, \hat{\rho}_2^{\mathrm{XFA}}) \equiv \Vert \hat{\rho}_2^{\mathrm{Full}} - \hat{\rho}_2^{\mathrm{XFA}} \Vert /2$ as a function of the coupling constant $\gamma$.
The trace distance for cat-state light irradiation (blue curve) scales as $\gamma^4$.
In contrast, for coherent-state irradiation, it scales as $\gamma^3$ and is always larger than that of cat-state light.
This demonstrates that, in the present system, our effective theory describes the electron dynamics more accurately for cat-state light than for the widely used laser light.

This scaling behavior can be qualitatively explained by photon dynamics influenced by nonequilibrium electrons.
In the next leading order of the Born approximation, we can formally consider the following equation for the photon field:
\begin{equation}
\mathrm{i} \partial_{\tau} \alpha(\tau) =  \omega \alpha(\tau) -\gamma \hat{\mathcal{J}}^-[\gamma \alpha_0(\tau)],\ \ \ \hat{\mathcal{J}}^- \equiv -\updelta \hat{\mathcal{H}}_{\mathrm{e}} / \updelta \gamma \overline{\alpha}. \label{eq:1st-order photon EoM}
\end{equation}
Here, $\hat{\mathcal{J}}^-$ is the electron current operator with positive frequency components, and together with the negative-frequency operator $\hat{\mathcal{J}}^+ \equiv -\updelta \hat{\mathcal{H}}_{\mathrm{e}}/\updelta \gamma \alpha$, they constitute the current operator $\hat{\mathcal{J}} \equiv -\updelta \hat{\mathcal{H}}_{\mathrm{e}}/\updelta A = \hat{\mathcal{J}}^+ + \hat{\mathcal{J}}^-$.
Here, $A = \gamma (\alpha + \overline{\alpha})$ denotes the vector potential.
Equation~\eqref{eq:1st-order photon EoM} describes the conventional electromagnetic response, where the photon field amplitude increases or decreases as an electric current flows.
The laser irradiation induces a finite electric current $\langle \hat{\mathcal{J}} \rangle \propto \gamma$, resulting in the vector potential modulation $\updelta A \sim \gamma^2 \langle \hat{\mathcal{J}} \rangle \sim \gamma^3$.
In contrast, under cat-state light irradiation, the expectation value of the current is zero, but the current fluctuation $\langle \hat{\mathcal{J}} \hat{\mathcal{J}} \rangle$ remains finite, leading to an interaction energy change proportional to $\gamma^2 \langle \hat{\mathcal{J}} \hat{\mathcal{J}} \rangle \propto \gamma^4$ via the photon-mediated electron--electron interactions (details are provided in the Supplemental Material).
This estimate indicates that the external-field approximation [Eqs.~\eqref{eq:P weighted density matrix} and~\eqref{eq:EoM}] remains valid as long as the electron--photon coupling constant $\gamma$ is small and the induced electric current and current fluctuations are small.

\textit{Outlook}---%
Our effective theory opens two promising avenues for future research.
The first focuses on the broad range of quantum light states.
Exploring other quantum superposed states, such as the bright squeezed vacuum state~\cite{Spasibko2017a, Manceau2019} and the Gottesman--Kitaev--Preskill state~\cite{Gottesman2001}, is critical for uncovering diverse physical phenomena arising from interferential non-Hermitian driving.
The second explores the applicability of the theory to various electron systems.
By treating quantum photons as an external field, the effective theory significantly reduces computational costs, enabling detailed analyses of condensed matter systems driven by quantum photons.
This approach paves the way for opportunities in quantum phase engineering with quantum light.

In addition, we have confirmed that the dynamics described by the equations of motion [Eqs.~\eqref{eq:P weighted density matrix} and~\eqref{eq:EoM}] preserve both the trace and positivity of the electron density matrix in the weak-coupling and weak-excitation regimes.
It is therefore essential to theoretically investigate cases where these approximations break down, as the positivity may no longer be preserved in such regimes.
This underscores the critical role of the coupled dynamics between electrons and quantum-superposed light. Addressing this issue by developing equations that ensure completely positive mappings~\cite{DAbbruzzo2024} remains a fundamentally important challenge for future work.

\begin{acknowledgments}
The authors thank Kunio Ishida, Naomichi Hatano, Kazuaki Takasan, and Taiga Nakamoto for fruitful discussions.
This work was supported by JST FOREST (Grant No.~JPMJFR2131) and JSPS KAKENHI (Grants No.~JP23K19030, No.~JP23K13052, No.~JP23K25805, No.~JP24H00191, and No.~JP24K00563).
\end{acknowledgments}

\bibliography{ref}

\begin{thebibliography}{65}%
\makeatletter
\providecommand \@ifxundefined [1]{%
 \@ifx{#1\undefined}
}%
\providecommand \@ifnum [1]{%
 \ifnum #1\expandafter \@firstoftwo
 \else \expandafter \@secondoftwo
 \fi
}%
\providecommand \@ifx [1]{%
 \ifx #1\expandafter \@firstoftwo
 \else \expandafter \@secondoftwo
 \fi
}%
\providecommand \natexlab [1]{#1}%
\providecommand \enquote  [1]{``#1''}%
\providecommand \bibnamefont  [1]{#1}%
\providecommand \bibfnamefont [1]{#1}%
\providecommand \citenamefont [1]{#1}%
\providecommand \href@noop [0]{\@secondoftwo}%
\providecommand \href [0]{\begingroup \@sanitize@url \@href}%
\providecommand \@href[1]{\@@startlink{#1}\@@href}%
\providecommand \@@href[1]{\endgroup#1\@@endlink}%
\providecommand \@sanitize@url [0]{\catcode `\\12\catcode `\$12\catcode `\&12\catcode `\#12\catcode `\^12\catcode `\_12\catcode `\%12\relax}%
\providecommand \@@startlink[1]{}%
\providecommand \@@endlink[0]{}%
\providecommand \url  [0]{\begingroup\@sanitize@url \@url }%
\providecommand \@url [1]{\endgroup\@href {#1}{\urlprefix }}%
\providecommand \urlprefix  [0]{URL }%
\providecommand \Eprint [0]{\href }%
\providecommand \doibase [0]{https://doi.org/}%
\providecommand \selectlanguage [0]{\@gobble}%
\providecommand \bibinfo  [0]{\@secondoftwo}%
\providecommand \bibfield  [0]{\@secondoftwo}%
\providecommand \translation [1]{[#1]}%
\providecommand \BibitemOpen [0]{}%
\providecommand \bibitemStop [0]{}%
\providecommand \bibitemNoStop [0]{.\EOS\space}%
\providecommand \EOS [0]{\spacefactor3000\relax}%
\providecommand \BibitemShut  [1]{\csname bibitem#1\endcsname}%
\let\auto@bib@innerbib\@empty
\bibitem [{\citenamefont {Schr\"{o}dinger}(1935)}]{Schrodinger1935}%
  \BibitemOpen
  \bibfield  {author} {\bibinfo {author} {\bibfnamefont {E.}~\bibnamefont {Schr\"{o}dinger}},\ }\bibfield  {title} {\bibinfo {title} {{Die gegenwärtige Situation in der Quantenmechanik}},\ }\href {https://doi.org/10.1007/BF01491891} {\bibfield  {journal} {\bibinfo  {journal} {Naturwissenschaften}\ }\textbf {\bibinfo {volume} {23}},\ \bibinfo {pages} {807} (\bibinfo {year} {1935})}\BibitemShut {NoStop}%
\bibitem [{\citenamefont {Wineland}(2013)}]{Wineland2013}%
  \BibitemOpen
  \bibfield  {author} {\bibinfo {author} {\bibfnamefont {D.~J.}\ \bibnamefont {Wineland}},\ }\bibfield  {title} {\bibinfo {title} {{Nobel Lecture: Superposition, entanglement, and raising Schr{\"{o}}dinger's cat}},\ }\href {https://doi.org/10.1103/RevModPhys.85.1103} {\bibfield  {journal} {\bibinfo  {journal} {Rev. Mod. Phys.}\ }\textbf {\bibinfo {volume} {85}},\ \bibinfo {pages} {1103} (\bibinfo {year} {2013})}\BibitemShut {NoStop}%
\bibitem [{\citenamefont {Fr{\"{o}}wis}\ \emph {et~al.}(2018)\citenamefont {Fr{\"{o}}wis}, \citenamefont {Sekatski}, \citenamefont {D{\"{u}}r}, \citenamefont {Gisin},\ and\ \citenamefont {Sangouard}}]{Frowis2018}%
  \BibitemOpen
  \bibfield  {author} {\bibinfo {author} {\bibfnamefont {F.}~\bibnamefont {Fr{\"{o}}wis}}, \bibinfo {author} {\bibfnamefont {P.}~\bibnamefont {Sekatski}}, \bibinfo {author} {\bibfnamefont {W.}~\bibnamefont {D{\"{u}}r}}, \bibinfo {author} {\bibfnamefont {N.}~\bibnamefont {Gisin}},\ and\ \bibinfo {author} {\bibfnamefont {N.}~\bibnamefont {Sangouard}},\ }\bibfield  {title} {\bibinfo {title} {{Macroscopic quantum states: Measures, fragility, and implementations}},\ }\href {https://doi.org/10.1103/RevModPhys.90.025004} {\bibfield  {journal} {\bibinfo  {journal} {Rev. Mod. Phys.}\ }\textbf {\bibinfo {volume} {90}},\ \bibinfo {pages} {025004} (\bibinfo {year} {2018})}\BibitemShut {NoStop}%
\bibitem [{\citenamefont {Shimizu}\ \emph {et~al.}(2013)\citenamefont {Shimizu}, \citenamefont {Matsuzaki},\ and\ \citenamefont {Ukena}}]{Shimizu2013}%
  \BibitemOpen
  \bibfield  {author} {\bibinfo {author} {\bibfnamefont {A.}~\bibnamefont {Shimizu}}, \bibinfo {author} {\bibfnamefont {Y.}~\bibnamefont {Matsuzaki}},\ and\ \bibinfo {author} {\bibfnamefont {A.}~\bibnamefont {Ukena}},\ }\bibfield  {title} {\bibinfo {title} {{Necessity of Superposition of Macroscopically Distinct States for Quantum Computational Speedup}},\ }\href {https://doi.org/10.7566/JPSJ.82.054801} {\bibfield  {journal} {\bibinfo  {journal} {J. Phys. Soc. Jpn.}\ }\textbf {\bibinfo {volume} {82}},\ \bibinfo {pages} {054801} (\bibinfo {year} {2013})}\BibitemShut {NoStop}%
\bibitem [{\citenamefont {Aberg}()}]{Aberg2006}%
  \BibitemOpen
  \bibfield  {author} {\bibinfo {author} {\bibfnamefont {J.}~\bibnamefont {Aberg}},\ }\bibfield  {title} {\bibinfo {title} {{Quantifying Superposition}},\ }\Eprint {https://arxiv.org/abs/quant-ph/0612146} {arXiv:quant-ph/0612146} \BibitemShut {NoStop}%
\bibitem [{\citenamefont {Stahlke}(2014)}]{Stahlke2014}%
  \BibitemOpen
  \bibfield  {author} {\bibinfo {author} {\bibfnamefont {D.}~\bibnamefont {Stahlke}},\ }\bibfield  {title} {\bibinfo {title} {{Quantum interference as a resource for quantum speedup}},\ }\href {https://doi.org/10.1103/PhysRevA.90.022302} {\bibfield  {journal} {\bibinfo  {journal} {Phys. Rev. A}\ }\textbf {\bibinfo {volume} {90}},\ \bibinfo {pages} {022302} (\bibinfo {year} {2014})}\BibitemShut {NoStop}%
\bibitem [{\citenamefont {Theurer}\ \emph {et~al.}(2017)\citenamefont {Theurer}, \citenamefont {Killoran}, \citenamefont {Egloff},\ and\ \citenamefont {Plenio}}]{Theurer2017}%
  \BibitemOpen
  \bibfield  {author} {\bibinfo {author} {\bibfnamefont {T.}~\bibnamefont {Theurer}}, \bibinfo {author} {\bibfnamefont {N.}~\bibnamefont {Killoran}}, \bibinfo {author} {\bibfnamefont {D.}~\bibnamefont {Egloff}},\ and\ \bibinfo {author} {\bibfnamefont {M.~B.}\ \bibnamefont {Plenio}},\ }\bibfield  {title} {\bibinfo {title} {{Resource Theory of Superposition}},\ }\href {https://doi.org/10.1103/PhysRevLett.119.230401} {\bibfield  {journal} {\bibinfo  {journal} {Phys. Rev. Lett.}\ }\textbf {\bibinfo {volume} {119}},\ \bibinfo {pages} {230401} (\bibinfo {year} {2017})}\BibitemShut {NoStop}%
\bibitem [{\citenamefont {Chitambar}\ and\ \citenamefont {Gour}(2019)}]{Chitambar2019}%
  \BibitemOpen
  \bibfield  {author} {\bibinfo {author} {\bibfnamefont {E.}~\bibnamefont {Chitambar}}\ and\ \bibinfo {author} {\bibfnamefont {G.}~\bibnamefont {Gour}},\ }\bibfield  {title} {\bibinfo {title} {{Quantum resource theories}},\ }\href {https://doi.org/10.1103/RevModPhys.91.025001} {\bibfield  {journal} {\bibinfo  {journal} {Rev. Mod. Phys.}\ }\textbf {\bibinfo {volume} {91}},\ \bibinfo {pages} {025001} (\bibinfo {year} {2019})}\BibitemShut {NoStop}%
\bibitem [{\citenamefont {Lewenstein}\ \emph {et~al.}(2021)\citenamefont {Lewenstein}, \citenamefont {Ciappina}, \citenamefont {Pisanty}, \citenamefont {Rivera-Dean}, \citenamefont {Stammer}, \citenamefont {Lamprou},\ and\ \citenamefont {Tzallas}}]{Lewenstein2021}%
  \BibitemOpen
  \bibfield  {author} {\bibinfo {author} {\bibfnamefont {M.}~\bibnamefont {Lewenstein}}, \bibinfo {author} {\bibfnamefont {M.~F.}\ \bibnamefont {Ciappina}}, \bibinfo {author} {\bibfnamefont {E.}~\bibnamefont {Pisanty}}, \bibinfo {author} {\bibfnamefont {J.}~\bibnamefont {Rivera-Dean}}, \bibinfo {author} {\bibfnamefont {P.}~\bibnamefont {Stammer}}, \bibinfo {author} {\bibfnamefont {T.}~\bibnamefont {Lamprou}},\ and\ \bibinfo {author} {\bibfnamefont {P.}~\bibnamefont {Tzallas}},\ }\bibfield  {title} {\bibinfo {title} {{Generation of optical Schr{\"{o}}dinger cat states in intense laser–matter interactions}},\ }\href {https://doi.org/10.1038/s41567-021-01317-w} {\bibfield  {journal} {\bibinfo  {journal} {Nat. Phys.}\ }\textbf {\bibinfo {volume} {17}},\ \bibinfo {pages} {1104} (\bibinfo {year} {2021})}\BibitemShut {NoStop}%
\bibitem [{\citenamefont {Rivera-Dean}\ \emph {et~al.}(2021)\citenamefont {Rivera-Dean}, \citenamefont {Stammer}, \citenamefont {Pisanty}, \citenamefont {Lamprou}, \citenamefont {Tzallas}, \citenamefont {Lewenstein},\ and\ \citenamefont {Ciappina}}]{Rivera-Dean2021}%
  \BibitemOpen
  \bibfield  {author} {\bibinfo {author} {\bibfnamefont {J.}~\bibnamefont {Rivera-Dean}}, \bibinfo {author} {\bibfnamefont {P.}~\bibnamefont {Stammer}}, \bibinfo {author} {\bibfnamefont {E.}~\bibnamefont {Pisanty}}, \bibinfo {author} {\bibfnamefont {T.}~\bibnamefont {Lamprou}}, \bibinfo {author} {\bibfnamefont {P.}~\bibnamefont {Tzallas}}, \bibinfo {author} {\bibfnamefont {M.}~\bibnamefont {Lewenstein}},\ and\ \bibinfo {author} {\bibfnamefont {M.~F.}\ \bibnamefont {Ciappina}},\ }\bibfield  {title} {\bibinfo {title} {{New schemes for creating large optical Schr{\"{o}}dinger cat states using strong laser fields}},\ }\href {https://doi.org/10.1007/s10825-021-01789-2} {\bibfield  {journal} {\bibinfo  {journal} {J. Comput. Electron.}\ }\textbf {\bibinfo {volume} {20}},\ \bibinfo {pages} {2111} (\bibinfo {year} {2021})}\BibitemShut {NoStop}%
\bibitem [{\citenamefont {Stammer}\ \emph {et~al.}(2023)\citenamefont {Stammer}, \citenamefont {Rivera-Dean}, \citenamefont {Maxwell}, \citenamefont {Lamprou}, \citenamefont {Ord{\'{o}}{\~{n}}ez}, \citenamefont {Ciappina}, \citenamefont {Tzallas},\ and\ \citenamefont {Lewenstein}}]{Stammer2023}%
  \BibitemOpen
  \bibfield  {author} {\bibinfo {author} {\bibfnamefont {P.}~\bibnamefont {Stammer}}, \bibinfo {author} {\bibfnamefont {J.}~\bibnamefont {Rivera-Dean}}, \bibinfo {author} {\bibfnamefont {A.}~\bibnamefont {Maxwell}}, \bibinfo {author} {\bibfnamefont {T.}~\bibnamefont {Lamprou}}, \bibinfo {author} {\bibfnamefont {A.}~\bibnamefont {Ord{\'{o}}{\~{n}}ez}}, \bibinfo {author} {\bibfnamefont {M.~F.}\ \bibnamefont {Ciappina}}, \bibinfo {author} {\bibfnamefont {P.}~\bibnamefont {Tzallas}},\ and\ \bibinfo {author} {\bibfnamefont {M.}~\bibnamefont {Lewenstein}},\ }\bibfield  {title} {\bibinfo {title} {{Quantum Electrodynamics of Intense Laser-Matter Interactions: A Tool for Quantum State Engineering}},\ }\href {https://doi.org/10.1103/PRXQuantum.4.010201} {\bibfield  {journal} {\bibinfo  {journal} {PRX Quantum}\ }\textbf {\bibinfo {volume} {4}},\ \bibinfo {pages} {010201} (\bibinfo {year} {2023})}\BibitemShut {NoStop}%
\bibitem [{\citenamefont {Cruz-Rodriguez}\ \emph {et~al.}(2024)\citenamefont {Cruz-Rodriguez}, \citenamefont {Dey}, \citenamefont {Freibert},\ and\ \citenamefont {Stammer}}]{Cruz-rodriguez2024}%
  \BibitemOpen
  \bibfield  {author} {\bibinfo {author} {\bibfnamefont {L.}~\bibnamefont {Cruz-Rodriguez}}, \bibinfo {author} {\bibfnamefont {D.}~\bibnamefont {Dey}}, \bibinfo {author} {\bibfnamefont {A.}~\bibnamefont {Freibert}},\ and\ \bibinfo {author} {\bibfnamefont {P.}~\bibnamefont {Stammer}},\ }\bibfield  {title} {\bibinfo {title} {{Quantum phenomena in attosecond science}},\ }\href {https://doi.org/10.1038/s42254-024-00769-2} {\bibfield  {journal} {\bibinfo  {journal} {Nat. Rev. Phys.}\ }\textbf {\bibinfo {volume} {6}},\ \bibinfo {pages} {691} (\bibinfo {year} {2024})}\BibitemShut {NoStop}%
\bibitem [{\citenamefont {Lamprou}\ \emph {et~al.}(2025{\natexlab{a}})\citenamefont {Lamprou}, \citenamefont {Stammer}, \citenamefont {Rivera-Dean}, \citenamefont {Tsatrafyllis}, \citenamefont {Ciappina}, \citenamefont {Lewenstein},\ and\ \citenamefont {Tzallas}}]{Lamprou_2025}%
  \BibitemOpen
  \bibfield  {author} {\bibinfo {author} {\bibfnamefont {T.}~\bibnamefont {Lamprou}}, \bibinfo {author} {\bibfnamefont {P.}~\bibnamefont {Stammer}}, \bibinfo {author} {\bibfnamefont {J.}~\bibnamefont {Rivera-Dean}}, \bibinfo {author} {\bibfnamefont {N.}~\bibnamefont {Tsatrafyllis}}, \bibinfo {author} {\bibfnamefont {M.~F.}\ \bibnamefont {Ciappina}}, \bibinfo {author} {\bibfnamefont {M.}~\bibnamefont {Lewenstein}},\ and\ \bibinfo {author} {\bibfnamefont {P.}~\bibnamefont {Tzallas}},\ }\bibfield  {title} {\bibinfo {title} {Recent developments in the generation of non-classical and entangled light states using intense laser-matter interactions},\ }\href {https://doi.org/10.1088/1361-6455/add9fe} {\bibfield  {journal} {\bibinfo  {journal} {Journal of Physics B: Atomic, Molecular and Optical Physics}\ }\textbf {\bibinfo {volume} {58}},\ \bibinfo {pages} {132001} (\bibinfo {year} {2025}{\natexlab{a}})}\BibitemShut {NoStop}%
\bibitem [{\citenamefont {Gonoskov}\ \emph {et~al.}(2024)\citenamefont {Gonoskov}, \citenamefont {Sondenheimer}, \citenamefont {H{\"{u}}necke}, \citenamefont {Kartashov}, \citenamefont {Peschel},\ and\ \citenamefont {Gr{\"{a}}fe}}]{Gonoskov2024}%
  \BibitemOpen
  \bibfield  {author} {\bibinfo {author} {\bibfnamefont {I.}~\bibnamefont {Gonoskov}}, \bibinfo {author} {\bibfnamefont {R.}~\bibnamefont {Sondenheimer}}, \bibinfo {author} {\bibfnamefont {C.}~\bibnamefont {H{\"{u}}necke}}, \bibinfo {author} {\bibfnamefont {D.}~\bibnamefont {Kartashov}}, \bibinfo {author} {\bibfnamefont {U.}~\bibnamefont {Peschel}},\ and\ \bibinfo {author} {\bibfnamefont {S.}~\bibnamefont {Gr{\"{a}}fe}},\ }\bibfield  {title} {\bibinfo {title} {{Nonclassical light generation and control from laser-driven semiconductor intraband excitations}},\ }\href {https://doi.org/10.1103/PhysRevB.109.125110} {\bibfield  {journal} {\bibinfo  {journal} {Phys. Rev. B}\ }\textbf {\bibinfo {volume} {109}},\ \bibinfo {pages} {125110} (\bibinfo {year} {2024})}\BibitemShut {NoStop}%
\bibitem [{\citenamefont {Dahan}\ \emph {et~al.}(2023)\citenamefont {Dahan}, \citenamefont {Baranes}, \citenamefont {Gorlach}, \citenamefont {Ruimy}, \citenamefont {Rivera},\ and\ \citenamefont {Kaminer}}]{Dahan2023}%
  \BibitemOpen
  \bibfield  {author} {\bibinfo {author} {\bibfnamefont {R.}~\bibnamefont {Dahan}}, \bibinfo {author} {\bibfnamefont {G.}~\bibnamefont {Baranes}}, \bibinfo {author} {\bibfnamefont {A.}~\bibnamefont {Gorlach}}, \bibinfo {author} {\bibfnamefont {R.}~\bibnamefont {Ruimy}}, \bibinfo {author} {\bibfnamefont {N.}~\bibnamefont {Rivera}},\ and\ \bibinfo {author} {\bibfnamefont {I.}~\bibnamefont {Kaminer}},\ }\bibfield  {title} {\bibinfo {title} {{Creation of Optical Cat and GKP States Using Shaped Free Electrons}},\ }\href {https://doi.org/10.1103/PhysRevX.13.031001} {\bibfield  {journal} {\bibinfo  {journal} {Phys. Rev. X}\ }\textbf {\bibinfo {volume} {13}},\ \bibinfo {pages} {031001} (\bibinfo {year} {2023})}\BibitemShut {NoStop}%
\bibitem [{\citenamefont {Konno}\ \emph {et~al.}(2024)\citenamefont {Konno}, \citenamefont {Asavanant}, \citenamefont {Hanamura}, \citenamefont {Nagayoshi}, \citenamefont {Fukui}, \citenamefont {Sakaguchi}, \citenamefont {Ide}, \citenamefont {China}, \citenamefont {Yabuno}, \citenamefont {Miki}, \citenamefont {Terai}, \citenamefont {Takase}, \citenamefont {Endo}, \citenamefont {Marek}, \citenamefont {Filip}, \citenamefont {van Loock},\ and\ \citenamefont {Furusawa}}]{Konno2024a}%
  \BibitemOpen
  \bibfield  {author} {\bibinfo {author} {\bibfnamefont {S.}~\bibnamefont {Konno}}, \bibinfo {author} {\bibfnamefont {W.}~\bibnamefont {Asavanant}}, \bibinfo {author} {\bibfnamefont {F.}~\bibnamefont {Hanamura}}, \bibinfo {author} {\bibfnamefont {H.}~\bibnamefont {Nagayoshi}}, \bibinfo {author} {\bibfnamefont {K.}~\bibnamefont {Fukui}}, \bibinfo {author} {\bibfnamefont {A.}~\bibnamefont {Sakaguchi}}, \bibinfo {author} {\bibfnamefont {R.}~\bibnamefont {Ide}}, \bibinfo {author} {\bibfnamefont {F.}~\bibnamefont {China}}, \bibinfo {author} {\bibfnamefont {M.}~\bibnamefont {Yabuno}}, \bibinfo {author} {\bibfnamefont {S.}~\bibnamefont {Miki}}, \bibinfo {author} {\bibfnamefont {H.}~\bibnamefont {Terai}}, \bibinfo {author} {\bibfnamefont {K.}~\bibnamefont {Takase}}, \bibinfo {author} {\bibfnamefont {M.}~\bibnamefont {Endo}}, \bibinfo {author} {\bibfnamefont {P.}~\bibnamefont {Marek}}, \bibinfo {author} {\bibfnamefont {R.}~\bibnamefont {Filip}}, \bibinfo {author} {\bibfnamefont {P.}~\bibnamefont {van Loock}},\ and\
  \bibinfo {author} {\bibfnamefont {A.}~\bibnamefont {Furusawa}},\ }\bibfield  {title} {\bibinfo {title} {{Logical states for fault-tolerant quantum computation with propagating light}},\ }\href {https://doi.org/10.1126/science.adk7560} {\bibfield  {journal} {\bibinfo  {journal} {Science}\ }\textbf {\bibinfo {volume} {383}},\ \bibinfo {pages} {289} (\bibinfo {year} {2024})}\BibitemShut {NoStop}%
\bibitem [{\citenamefont {Stammer}\ \emph {et~al.}(2024)\citenamefont {Stammer}, \citenamefont {{Fern{\'{a}}ndez Martos}}, \citenamefont {Lewenstein},\ and\ \citenamefont {Rajchel-Mieldzio{\'{c}}}}]{Stammer2024b}%
  \BibitemOpen
  \bibfield  {author} {\bibinfo {author} {\bibfnamefont {P.}~\bibnamefont {Stammer}}, \bibinfo {author} {\bibfnamefont {T.}~\bibnamefont {{Fern{\'{a}}ndez Martos}}}, \bibinfo {author} {\bibfnamefont {M.}~\bibnamefont {Lewenstein}},\ and\ \bibinfo {author} {\bibfnamefont {G.}~\bibnamefont {Rajchel-Mieldzio{\'{c}}}},\ }\bibfield  {title} {\bibinfo {title} {{Metrological robustness of high photon number optical cat states}},\ }\href {https://doi.org/10.1088/2058-9565/ad7881} {\bibfield  {journal} {\bibinfo  {journal} {Quantum Sci. Technol.}\ }\textbf {\bibinfo {volume} {9}},\ \bibinfo {pages} {045047} (\bibinfo {year} {2024})}\BibitemShut {NoStop}%
\bibitem [{\citenamefont {Kira}\ \emph {et~al.}(2011)\citenamefont {Kira}, \citenamefont {Koch}, \citenamefont {Smith}, \citenamefont {Hunter},\ and\ \citenamefont {Cundiff}}]{Kira2011}%
  \BibitemOpen
  \bibfield  {author} {\bibinfo {author} {\bibfnamefont {M.}~\bibnamefont {Kira}}, \bibinfo {author} {\bibfnamefont {S.~W.}\ \bibnamefont {Koch}}, \bibinfo {author} {\bibfnamefont {R.~P.}\ \bibnamefont {Smith}}, \bibinfo {author} {\bibfnamefont {A.~E.}\ \bibnamefont {Hunter}},\ and\ \bibinfo {author} {\bibfnamefont {S.~T.}\ \bibnamefont {Cundiff}},\ }\bibfield  {title} {\bibinfo {title} {{Quantum spectroscopy with Schr{\"{o}}dinger-cat states}},\ }\href {https://doi.org/10.1038/nphys2091} {\bibfield  {journal} {\bibinfo  {journal} {Nat. Phys.}\ }\textbf {\bibinfo {volume} {7}},\ \bibinfo {pages} {799} (\bibinfo {year} {2011})}\BibitemShut {NoStop}%
\bibitem [{\citenamefont {Lamprou}\ \emph {et~al.}(2025{\natexlab{b}})\citenamefont {Lamprou}, \citenamefont {Rivera-Dean}, \citenamefont {Stammer}, \citenamefont {Lewenstein},\ and\ \citenamefont {Tzallas}}]{Lamprou2025}%
  \BibitemOpen
  \bibfield  {author} {\bibinfo {author} {\bibfnamefont {T.}~\bibnamefont {Lamprou}}, \bibinfo {author} {\bibfnamefont {J.}~\bibnamefont {Rivera-Dean}}, \bibinfo {author} {\bibfnamefont {P.}~\bibnamefont {Stammer}}, \bibinfo {author} {\bibfnamefont {M.}~\bibnamefont {Lewenstein}},\ and\ \bibinfo {author} {\bibfnamefont {P.}~\bibnamefont {Tzallas}},\ }\bibfield  {title} {\bibinfo {title} {{Nonlinear Optics Using Intense Optical Coherent State Superpositions}},\ }\href {https://doi.org/10.1103/PhysRevLett.134.013601} {\bibfield  {journal} {\bibinfo  {journal} {Phys. Rev. Lett.}\ }\textbf {\bibinfo {volume} {134}},\ \bibinfo {pages} {013601} (\bibinfo {year} {2025}{\natexlab{b}})}\BibitemShut {NoStop}%
\bibitem [{\citenamefont {Dorfman}\ \emph {et~al.}(2016)\citenamefont {Dorfman}, \citenamefont {Schlawin},\ and\ \citenamefont {Mukamel}}]{Dorfman2016}%
  \BibitemOpen
  \bibfield  {author} {\bibinfo {author} {\bibfnamefont {K.~E.}\ \bibnamefont {Dorfman}}, \bibinfo {author} {\bibfnamefont {F.}~\bibnamefont {Schlawin}},\ and\ \bibinfo {author} {\bibfnamefont {S.}~\bibnamefont {Mukamel}},\ }\bibfield  {title} {\bibinfo {title} {{Nonlinear optical signals and spectroscopy with quantum light}},\ }\href {https://doi.org/10.1103/RevModPhys.88.045008} {\bibfield  {journal} {\bibinfo  {journal} {Rev. Mod. Phys.}\ }\textbf {\bibinfo {volume} {88}},\ \bibinfo {pages} {045008} (\bibinfo {year} {2016})}\BibitemShut {NoStop}%
\bibitem [{\citenamefont {Mukamel}\ \emph {et~al.}(2020)\citenamefont {Mukamel}, \citenamefont {Freyberger}, \citenamefont {Schleich}, \citenamefont {Bellini}, \citenamefont {Zavatta}, \citenamefont {Leuchs}, \citenamefont {Silberhorn}, \citenamefont {Boyd}, \citenamefont {S{\'{a}}nchez-Soto}, \citenamefont {Stefanov}, \citenamefont {Barbieri}, \citenamefont {Paterova}, \citenamefont {Krivitsky}, \citenamefont {Shwartz}, \citenamefont {Tamasaku}, \citenamefont {Dorfman}, \citenamefont {Schlawin}, \citenamefont {Sandoghdar}, \citenamefont {Raymer}, \citenamefont {Marcus}, \citenamefont {Varnavski}, \citenamefont {Goodson}, \citenamefont {Zhou}, \citenamefont {Shi}, \citenamefont {Asban}, \citenamefont {Scully}, \citenamefont {Agarwal}, \citenamefont {Peng}, \citenamefont {Sokolov}, \citenamefont {Zhang}, \citenamefont {Zubairy}, \citenamefont {Vartanyants}, \citenamefont {del Valle},\ and\ \citenamefont {Laussy}}]{Mukamel2017}%
  \BibitemOpen
  \bibfield  {author} {\bibinfo {author} {\bibfnamefont {S.}~\bibnamefont {Mukamel}}, \bibinfo {author} {\bibfnamefont {M.}~\bibnamefont {Freyberger}}, \bibinfo {author} {\bibfnamefont {W.}~\bibnamefont {Schleich}}, \bibinfo {author} {\bibfnamefont {M.}~\bibnamefont {Bellini}}, \bibinfo {author} {\bibfnamefont {A.}~\bibnamefont {Zavatta}}, \bibinfo {author} {\bibfnamefont {G.}~\bibnamefont {Leuchs}}, \bibinfo {author} {\bibfnamefont {C.}~\bibnamefont {Silberhorn}}, \bibinfo {author} {\bibfnamefont {R.~W.}\ \bibnamefont {Boyd}}, \bibinfo {author} {\bibfnamefont {L.~L.}\ \bibnamefont {S{\'{a}}nchez-Soto}}, \bibinfo {author} {\bibfnamefont {A.}~\bibnamefont {Stefanov}}, \bibinfo {author} {\bibfnamefont {M.}~\bibnamefont {Barbieri}}, \bibinfo {author} {\bibfnamefont {A.}~\bibnamefont {Paterova}}, \bibinfo {author} {\bibfnamefont {L.}~\bibnamefont {Krivitsky}}, \bibinfo {author} {\bibfnamefont {S.}~\bibnamefont {Shwartz}}, \bibinfo {author} {\bibfnamefont {K.}~\bibnamefont {Tamasaku}}, \bibinfo {author}
  {\bibfnamefont {K.}~\bibnamefont {Dorfman}}, \bibinfo {author} {\bibfnamefont {F.}~\bibnamefont {Schlawin}}, \bibinfo {author} {\bibfnamefont {V.}~\bibnamefont {Sandoghdar}}, \bibinfo {author} {\bibfnamefont {M.}~\bibnamefont {Raymer}}, \bibinfo {author} {\bibfnamefont {A.}~\bibnamefont {Marcus}}, \bibinfo {author} {\bibfnamefont {O.}~\bibnamefont {Varnavski}}, \bibinfo {author} {\bibfnamefont {T.}~\bibnamefont {Goodson}}, \bibinfo {author} {\bibfnamefont {Z.-Y.}\ \bibnamefont {Zhou}}, \bibinfo {author} {\bibfnamefont {B.-S.}\ \bibnamefont {Shi}}, \bibinfo {author} {\bibfnamefont {S.}~\bibnamefont {Asban}}, \bibinfo {author} {\bibfnamefont {M.}~\bibnamefont {Scully}}, \bibinfo {author} {\bibfnamefont {G.}~\bibnamefont {Agarwal}}, \bibinfo {author} {\bibfnamefont {T.}~\bibnamefont {Peng}}, \bibinfo {author} {\bibfnamefont {A.~V.}\ \bibnamefont {Sokolov}}, \bibinfo {author} {\bibfnamefont {Z.-D.}\ \bibnamefont {Zhang}}, \bibinfo {author} {\bibfnamefont {M.~S.}\ \bibnamefont {Zubairy}}, \bibinfo {author}
  {\bibfnamefont {I.~A.}\ \bibnamefont {Vartanyants}}, \bibinfo {author} {\bibfnamefont {E.}~\bibnamefont {del Valle}},\ and\ \bibinfo {author} {\bibfnamefont {F.}~\bibnamefont {Laussy}},\ }\bibfield  {title} {\bibinfo {title} {{Roadmap on quantum light spectroscopy}},\ }\href {https://doi.org/10.1088/1361-6455/ab69a8} {\bibfield  {journal} {\bibinfo  {journal} {J. Phys. B At. Mol. Opt. Phys.}\ }\textbf {\bibinfo {volume} {53}},\ \bibinfo {pages} {072002} (\bibinfo {year} {2020})}\BibitemShut {NoStop}%
\bibitem [{\citenamefont {Lamprou}\ \emph {et~al.}(2020)\citenamefont {Lamprou}, \citenamefont {Liontos}, \citenamefont {Papadakis},\ and\ \citenamefont {Tzallas}}]{Lamprou2020}%
  \BibitemOpen
  \bibfield  {author} {\bibinfo {author} {\bibfnamefont {T.}~\bibnamefont {Lamprou}}, \bibinfo {author} {\bibfnamefont {I.}~\bibnamefont {Liontos}}, \bibinfo {author} {\bibfnamefont {N.~C.}\ \bibnamefont {Papadakis}},\ and\ \bibinfo {author} {\bibfnamefont {P.}~\bibnamefont {Tzallas}},\ }\bibfield  {title} {\bibinfo {title} {{A perspective on high photon flux nonclassical light and applications in nonlinear optics}},\ }\href {https://doi.org/10.1017/hpl.2020.44} {\bibfield  {journal} {\bibinfo  {journal} {High Power Laser Sci. Eng.}\ }\textbf {\bibinfo {volume} {8}},\ \bibinfo {pages} {e42} (\bibinfo {year} {2020})}\BibitemShut {NoStop}%
\bibitem [{\citenamefont {Lysne}\ and\ \citenamefont {Werner}(2021)}]{Lysne2021a}%
  \BibitemOpen
  \bibfield  {author} {\bibinfo {author} {\bibfnamefont {M.}~\bibnamefont {Lysne}}\ and\ \bibinfo {author} {\bibfnamefont {P.}~\bibnamefont {Werner}},\ }\bibfield  {title} {\bibinfo {title} {{Nonlinear optical processes in cavity light-matter systems}},\ }\href {https://doi.org/10.1103/PhysRevB.104.035209} {\bibfield  {journal} {\bibinfo  {journal} {Phys. Rev. B}\ }\textbf {\bibinfo {volume} {104}},\ \bibinfo {pages} {035209} (\bibinfo {year} {2021})}\BibitemShut {NoStop}%
\bibitem [{\citenamefont {Fedotov}\ \emph {et~al.}(2023)\citenamefont {Fedotov}, \citenamefont {Ilderton}, \citenamefont {Karbstein}, \citenamefont {King}, \citenamefont {Seipt}, \citenamefont {Taya},\ and\ \citenamefont {Torgrimsson}}]{Fedotov2023}%
  \BibitemOpen
  \bibfield  {author} {\bibinfo {author} {\bibfnamefont {A.}~\bibnamefont {Fedotov}}, \bibinfo {author} {\bibfnamefont {A.}~\bibnamefont {Ilderton}}, \bibinfo {author} {\bibfnamefont {F.}~\bibnamefont {Karbstein}}, \bibinfo {author} {\bibfnamefont {B.}~\bibnamefont {King}}, \bibinfo {author} {\bibfnamefont {D.}~\bibnamefont {Seipt}}, \bibinfo {author} {\bibfnamefont {H.}~\bibnamefont {Taya}},\ and\ \bibinfo {author} {\bibfnamefont {G.}~\bibnamefont {Torgrimsson}},\ }\bibfield  {title} {\bibinfo {title} {{Advances in QED with intense background fields}},\ }\href {https://doi.org/10.1016/j.physrep.2023.01.003} {\bibfield  {journal} {\bibinfo  {journal} {Phys. Rep.}\ }\textbf {\bibinfo {volume} {1010}},\ \bibinfo {pages} {1} (\bibinfo {year} {2023})}\BibitemShut {NoStop}%
\bibitem [{\citenamefont {Sennary}\ \emph {et~al.}(2025)\citenamefont {Sennary}, \citenamefont {Rivera-Dean}, \citenamefont {ElKabbash}, \citenamefont {Pervak}, \citenamefont {Lewenstein},\ and\ \citenamefont {Hassan}}]{Sennary2024}%
  \BibitemOpen
  \bibfield  {author} {\bibinfo {author} {\bibfnamefont {M.}~\bibnamefont {Sennary}}, \bibinfo {author} {\bibfnamefont {J.}~\bibnamefont {Rivera-Dean}}, \bibinfo {author} {\bibfnamefont {M.}~\bibnamefont {ElKabbash}}, \bibinfo {author} {\bibfnamefont {V.}~\bibnamefont {Pervak}}, \bibinfo {author} {\bibfnamefont {M.}~\bibnamefont {Lewenstein}},\ and\ \bibinfo {author} {\bibfnamefont {M.~T.}\ \bibnamefont {Hassan}},\ }\bibfield  {title} {\bibinfo {title} {{Attosecond quantum uncertainty dynamics and ultrafast squeezed light for quantum communication}},\ }\href {https://doi.org/10.1038/s41377-025-02055-x} {\bibfield  {journal} {\bibinfo  {journal} {Light Sci. Appl.}\ }\textbf {\bibinfo {volume} {14}},\ \bibinfo {pages} {350} (\bibinfo {year} {2025})}\BibitemShut {NoStop}%
\bibitem [{\citenamefont {Gorini}\ \emph {et~al.}(1976)\citenamefont {Gorini}, \citenamefont {Kossakowski},\ and\ \citenamefont {Sudarshan}}]{Gorini1975}%
  \BibitemOpen
  \bibfield  {author} {\bibinfo {author} {\bibfnamefont {V.}~\bibnamefont {Gorini}}, \bibinfo {author} {\bibfnamefont {A.}~\bibnamefont {Kossakowski}},\ and\ \bibinfo {author} {\bibfnamefont {E.~C.~G.}\ \bibnamefont {Sudarshan}},\ }\bibfield  {title} {\bibinfo {title} {{Completely positive dynamical semigroups of $N$-level systems}},\ }\href {https://doi.org/10.1063/1.522979} {\bibfield  {journal} {\bibinfo  {journal} {J. Math. Phys.}\ }\textbf {\bibinfo {volume} {17}},\ \bibinfo {pages} {821} (\bibinfo {year} {1976})}\BibitemShut {NoStop}%
\bibitem [{\citenamefont {Lindblad}(1976)}]{Lindblad1976}%
  \BibitemOpen
  \bibfield  {author} {\bibinfo {author} {\bibfnamefont {G.}~\bibnamefont {Lindblad}},\ }\bibfield  {title} {\bibinfo {title} {{On the generators of quantum dynamical semigroups}},\ }\href {https://doi.org/10.1007/BF01608499} {\bibfield  {journal} {\bibinfo  {journal} {Commun. Math. Phys.}\ }\textbf {\bibinfo {volume} {48}},\ \bibinfo {pages} {119} (\bibinfo {year} {1976})}\BibitemShut {NoStop}%
\bibitem [{\citenamefont {Agarwal}\ and\ \citenamefont {Puri}(1990)}]{Agarwal1990}%
  \BibitemOpen
  \bibfield  {author} {\bibinfo {author} {\bibfnamefont {G.~S.}\ \bibnamefont {Agarwal}}\ and\ \bibinfo {author} {\bibfnamefont {R.~R.}\ \bibnamefont {Puri}},\ }\bibfield  {title} {\bibinfo {title} {{Cooperative behavior of atoms irradiated by broadband squeezed light}},\ }\href {https://doi.org/10.1103/PhysRevA.41.3782} {\bibfield  {journal} {\bibinfo  {journal} {Phys. Rev. A}\ }\textbf {\bibinfo {volume} {41}},\ \bibinfo {pages} {3782} (\bibinfo {year} {1990})}\BibitemShut {NoStop}%
\bibitem [{\citenamefont {Gorlach}\ \emph {et~al.}(2023)\citenamefont {Gorlach}, \citenamefont {Tzur}, \citenamefont {Birk}, \citenamefont {Kr{\"{u}}ger}, \citenamefont {Rivera}, \citenamefont {Cohen},\ and\ \citenamefont {Kaminer}}]{Gorlach2023}%
  \BibitemOpen
  \bibfield  {author} {\bibinfo {author} {\bibfnamefont {A.}~\bibnamefont {Gorlach}}, \bibinfo {author} {\bibfnamefont {M.~E.}\ \bibnamefont {Tzur}}, \bibinfo {author} {\bibfnamefont {M.}~\bibnamefont {Birk}}, \bibinfo {author} {\bibfnamefont {M.}~\bibnamefont {Kr{\"{u}}ger}}, \bibinfo {author} {\bibfnamefont {N.}~\bibnamefont {Rivera}}, \bibinfo {author} {\bibfnamefont {O.}~\bibnamefont {Cohen}},\ and\ \bibinfo {author} {\bibfnamefont {I.}~\bibnamefont {Kaminer}},\ }\bibfield  {title} {\bibinfo {title} {{High-harmonic generation driven by quantum light}},\ }\href {https://doi.org/10.1038/s41567-023-02127-y} {\bibfield  {journal} {\bibinfo  {journal} {Nat. Phys.}\ }\textbf {\bibinfo {volume} {19}},\ \bibinfo {pages} {1689} (\bibinfo {year} {2023})}\BibitemShut {NoStop}%
\bibitem [{\citenamefont {{Even Tzur}}\ \emph {et~al.}(2023)\citenamefont {{Even Tzur}}, \citenamefont {Birk}, \citenamefont {Gorlach}, \citenamefont {Kr{\"{u}}ger}, \citenamefont {Kaminer},\ and\ \citenamefont {Cohen}}]{Tzur2022a}%
  \BibitemOpen
  \bibfield  {author} {\bibinfo {author} {\bibfnamefont {M.}~\bibnamefont {{Even Tzur}}}, \bibinfo {author} {\bibfnamefont {M.}~\bibnamefont {Birk}}, \bibinfo {author} {\bibfnamefont {A.}~\bibnamefont {Gorlach}}, \bibinfo {author} {\bibfnamefont {M.}~\bibnamefont {Kr{\"{u}}ger}}, \bibinfo {author} {\bibfnamefont {I.}~\bibnamefont {Kaminer}},\ and\ \bibinfo {author} {\bibfnamefont {O.}~\bibnamefont {Cohen}},\ }\bibfield  {title} {\bibinfo {title} {{Photon-statistics force in ultrafast electron dynamics}},\ }\href {https://doi.org/10.1038/s41566-023-01209-w} {\bibfield  {journal} {\bibinfo  {journal} {Nat. Photonics}\ }\textbf {\bibinfo {volume} {17}},\ \bibinfo {pages} {501} (\bibinfo {year} {2023})}\BibitemShut {NoStop}%
\bibitem [{\citenamefont {Khalaf}\ and\ \citenamefont {Kaminer}(2023)}]{Khalaf2023}%
  \BibitemOpen
  \bibfield  {author} {\bibinfo {author} {\bibfnamefont {M.}~\bibnamefont {Khalaf}}\ and\ \bibinfo {author} {\bibfnamefont {I.}~\bibnamefont {Kaminer}},\ }\bibfield  {title} {\bibinfo {title} {{Compton scattering driven by intense quantum light}},\ }\href {https://doi.org/10.1126/sciadv.ade0932} {\bibfield  {journal} {\bibinfo  {journal} {Sci. Adv.}\ }\textbf {\bibinfo {volume} {9}},\ \bibinfo {pages} {eade0932} (\bibinfo {year} {2023})}\BibitemShut {NoStop}%
\bibitem [{\citenamefont {Rasputnyi}\ \emph {et~al.}(2024)\citenamefont {Rasputnyi}, \citenamefont {Chen}, \citenamefont {Birk}, \citenamefont {Cohen}, \citenamefont {Kaminer}, \citenamefont {Kr{\"{u}}ger}, \citenamefont {Seletskiy}, \citenamefont {Chekhova},\ and\ \citenamefont {Tani}}]{Rasputnyi2024a}%
  \BibitemOpen
  \bibfield  {author} {\bibinfo {author} {\bibfnamefont {A.}~\bibnamefont {Rasputnyi}}, \bibinfo {author} {\bibfnamefont {Z.}~\bibnamefont {Chen}}, \bibinfo {author} {\bibfnamefont {M.}~\bibnamefont {Birk}}, \bibinfo {author} {\bibfnamefont {O.}~\bibnamefont {Cohen}}, \bibinfo {author} {\bibfnamefont {I.}~\bibnamefont {Kaminer}}, \bibinfo {author} {\bibfnamefont {M.}~\bibnamefont {Kr{\"{u}}ger}}, \bibinfo {author} {\bibfnamefont {D.}~\bibnamefont {Seletskiy}}, \bibinfo {author} {\bibfnamefont {M.}~\bibnamefont {Chekhova}},\ and\ \bibinfo {author} {\bibfnamefont {F.}~\bibnamefont {Tani}},\ }\bibfield  {title} {\bibinfo {title} {{High-harmonic generation by a bright squeezed vacuum}},\ }\href {https://doi.org/10.1038/s41567-024-02659-x} {\bibfield  {journal} {\bibinfo  {journal} {Nat. Phys.}\ }\textbf {\bibinfo {volume} {20}},\ \bibinfo {pages} {1960} (\bibinfo {year} {2024})}\BibitemShut {NoStop}%
\bibitem [{\citenamefont {{Even Tzur}}\ and\ \citenamefont {Cohen}(2024)}]{EvenTzur2024}%
  \BibitemOpen
  \bibfield  {author} {\bibinfo {author} {\bibfnamefont {M.}~\bibnamefont {{Even Tzur}}}\ and\ \bibinfo {author} {\bibfnamefont {O.}~\bibnamefont {Cohen}},\ }\bibfield  {title} {\bibinfo {title} {{Motion of charged particles in bright squeezed vacuum}},\ }\href {https://doi.org/10.1038/s41377-024-01381-w} {\bibfield  {journal} {\bibinfo  {journal} {Light Sci. Appl.}\ }\textbf {\bibinfo {volume} {13}},\ \bibinfo {pages} {41} (\bibinfo {year} {2024})}\BibitemShut {NoStop}%
\bibitem [{\citenamefont {Wang}\ \emph {et~al.}(2025)\citenamefont {Wang}, \citenamefont {Lai},\ and\ \citenamefont {Liu}}]{4x11-phs2}%
  \BibitemOpen
  \bibfield  {author} {\bibinfo {author} {\bibfnamefont {S.}~\bibnamefont {Wang}}, \bibinfo {author} {\bibfnamefont {X.}~\bibnamefont {Lai}},\ and\ \bibinfo {author} {\bibfnamefont {X.}~\bibnamefont {Liu}},\ }\bibfield  {title} {\bibinfo {title} {Attosecond pulse synthesis from high-order harmonic generation in intense squeezed light},\ }\href {https://doi.org/10.1103/4x11-phs2} {\bibfield  {journal} {\bibinfo  {journal} {Phys. Rev. A}\ }\textbf {\bibinfo {volume} {112}},\ \bibinfo {pages} {L011102} (\bibinfo {year} {2025})}\BibitemShut {NoStop}%
\bibitem [{\citenamefont {Sudarshan}(1963)}]{Sudarshan1963}%
  \BibitemOpen
  \bibfield  {author} {\bibinfo {author} {\bibfnamefont {E.~C.~G.}\ \bibnamefont {Sudarshan}},\ }\bibfield  {title} {\bibinfo {title} {{Equivalence of Semiclassical and Quantum Mechanical Descriptions of Statistical Light Beams}},\ }\href {https://doi.org/10.1103/PhysRevLett.10.277} {\bibfield  {journal} {\bibinfo  {journal} {Phys. Rev. Lett.}\ }\textbf {\bibinfo {volume} {10}},\ \bibinfo {pages} {277} (\bibinfo {year} {1963})}\BibitemShut {NoStop}%
\bibitem [{\citenamefont {Glauber}(1963)}]{Glauber1963a}%
  \BibitemOpen
  \bibfield  {author} {\bibinfo {author} {\bibfnamefont {R.~J.}\ \bibnamefont {Glauber}},\ }\bibfield  {title} {\bibinfo {title} {{Coherent and Incoherent States of the Radiation Field}},\ }\href {https://doi.org/10.1103/PhysRev.131.2766} {\bibfield  {journal} {\bibinfo  {journal} {Phys. Rev.}\ }\textbf {\bibinfo {volume} {131}},\ \bibinfo {pages} {2766} (\bibinfo {year} {1963})}\BibitemShut {NoStop}%
\bibitem [{\citenamefont {Feynman}\ \emph {et~al.}(2010)\citenamefont {Feynman}, \citenamefont {Hibbs},\ and\ \citenamefont {Styer}}]{Feynman:100771}%
  \BibitemOpen
  \bibfield  {author} {\bibinfo {author} {\bibfnamefont {R.~P.}\ \bibnamefont {Feynman}}, \bibinfo {author} {\bibfnamefont {A.~R.}\ \bibnamefont {Hibbs}},\ and\ \bibinfo {author} {\bibfnamefont {D.~F.}\ \bibnamefont {Styer}},\ }\href@noop {} {\emph {\bibinfo {title} {{Quantum Mechanics and Path Integrals: Emended Edition}}}}\ (\bibinfo  {publisher} {Dover Publications},\ \bibinfo {address} {New York, NY},\ \bibinfo {year} {2010})\BibitemShut {NoStop}%
\bibitem [{\citenamefont {Brewster}\ and\ \citenamefont {Franson}(2018)}]{Brewster2018}%
  \BibitemOpen
  \bibfield  {author} {\bibinfo {author} {\bibfnamefont {R.~A.}\ \bibnamefont {Brewster}}\ and\ \bibinfo {author} {\bibfnamefont {J.~D.}\ \bibnamefont {Franson}},\ }\bibfield  {title} {\bibinfo {title} {{Generalized delta functions and their use in quantum optics}},\ }\href {https://doi.org/10.1063/1.4985938} {\bibfield  {journal} {\bibinfo  {journal} {J. Math. Phys.}\ }\textbf {\bibinfo {volume} {59}},\ \bibinfo {pages} {012102} (\bibinfo {year} {2018})}\BibitemShut {NoStop}%
\bibitem [{\citenamefont {Hatano}\ and\ \citenamefont {Nelson}(1996)}]{Hatano1996}%
  \BibitemOpen
  \bibfield  {author} {\bibinfo {author} {\bibfnamefont {N.}~\bibnamefont {Hatano}}\ and\ \bibinfo {author} {\bibfnamefont {D.~R.}\ \bibnamefont {Nelson}},\ }\bibfield  {title} {\bibinfo {title} {{Localization Transitions in Non-Hermitian Quantum Mechanics}},\ }\href {https://doi.org/10.1103/PhysRevLett.77.570} {\bibfield  {journal} {\bibinfo  {journal} {Phys. Rev. Lett.}\ }\textbf {\bibinfo {volume} {77}},\ \bibinfo {pages} {570} (\bibinfo {year} {1996})}\BibitemShut {NoStop}%
\bibitem [{\citenamefont {Bender}\ and\ \citenamefont {Boettcher}(1998)}]{Bender1998}%
  \BibitemOpen
  \bibfield  {author} {\bibinfo {author} {\bibfnamefont {C.~M.}\ \bibnamefont {Bender}}\ and\ \bibinfo {author} {\bibfnamefont {S.}~\bibnamefont {Boettcher}},\ }\bibfield  {title} {\bibinfo {title} {Real spectra in non-hermitian hamiltonians having $\mathscr{P}\mathscr{T}$ symmetry},\ }\href {https://doi.org/10.1103/PhysRevLett.80.5243} {\bibfield  {journal} {\bibinfo  {journal} {Phys. Rev. Lett.}\ }\textbf {\bibinfo {volume} {80}},\ \bibinfo {pages} {5243} (\bibinfo {year} {1998})}\BibitemShut {NoStop}%
\bibitem [{\citenamefont {Brody}(2014)}]{Brody2014}%
  \BibitemOpen
  \bibfield  {author} {\bibinfo {author} {\bibfnamefont {D.~C.}\ \bibnamefont {Brody}},\ }\bibfield  {title} {\bibinfo {title} {{Biorthogonal quantum mechanics}},\ }\href {https://doi.org/10.1088/1751-8113/47/3/035305} {\bibfield  {journal} {\bibinfo  {journal} {J. Phys. A Math. Theor.}\ }\textbf {\bibinfo {volume} {47}},\ \bibinfo {pages} {035305} (\bibinfo {year} {2014})}\BibitemShut {NoStop}%
\bibitem [{\citenamefont {Sza{\'{n}}kowski}(2023)}]{Szankowski2023}%
  \BibitemOpen
  \bibfield  {author} {\bibinfo {author} {\bibfnamefont {P.}~\bibnamefont {Sza{\'{n}}kowski}},\ }\bibfield  {title} {\bibinfo {title} {{Introduction to the theory of open quantum systems}},\ }\href {https://doi.org/10.21468/SciPostPhysLectNotes.68} {\bibfield  {journal} {\bibinfo  {journal} {SciPost Phys. Lect. Notes}\ }\textbf {\bibinfo {volume} {68}},\ \bibinfo {pages} {68} (\bibinfo {year} {2023})}\BibitemShut {NoStop}%
\bibitem [{\citenamefont {Felski}\ \emph {et~al.}()\citenamefont {Felski}, \citenamefont {Beygi}, \citenamefont {Karapoulitidis},\ and\ \citenamefont {Klevansky}}]{Felski2024}%
  \BibitemOpen
  \bibfield  {author} {\bibinfo {author} {\bibfnamefont {A.}~\bibnamefont {Felski}}, \bibinfo {author} {\bibfnamefont {A.}~\bibnamefont {Beygi}}, \bibinfo {author} {\bibfnamefont {C.}~\bibnamefont {Karapoulitidis}},\ and\ \bibinfo {author} {\bibfnamefont {S.~P.}\ \bibnamefont {Klevansky}},\ }\bibfield  {title} {\bibinfo {title} {{Three perspectives on entropy dynamics in a non-Hermitian two-state system}},\ }\Eprint {https://arxiv.org/abs/2404.03492} {arXiv:2404.03492} \BibitemShut {NoStop}%
\bibitem [{\citenamefont {Dicke}(1954)}]{Dicke1954}%
  \BibitemOpen
  \bibfield  {author} {\bibinfo {author} {\bibfnamefont {R.~H.}\ \bibnamefont {Dicke}},\ }\bibfield  {title} {\bibinfo {title} {{Coherence in Spontaneous Radiation Processes}},\ }\href {https://doi.org/10.1103/PhysRev.93.99} {\bibfield  {journal} {\bibinfo  {journal} {Phys. Rev.}\ }\textbf {\bibinfo {volume} {93}},\ \bibinfo {pages} {99} (\bibinfo {year} {1954})}\BibitemShut {NoStop}%
\bibitem [{\citenamefont {Rabi}(1937)}]{Rabi1937}%
  \BibitemOpen
  \bibfield  {author} {\bibinfo {author} {\bibfnamefont {I.~I.}\ \bibnamefont {Rabi}},\ }\bibfield  {title} {\bibinfo {title} {{Space Quantization in a Gyrating Magnetic Field}},\ }\href {https://doi.org/10.1103/PhysRev.51.652} {\bibfield  {journal} {\bibinfo  {journal} {Phys. Rev.}\ }\textbf {\bibinfo {volume} {51}},\ \bibinfo {pages} {652} (\bibinfo {year} {1937})}\BibitemShut {NoStop}%
\bibitem [{\citenamefont {Hanamura}(1993)}]{Hanamura1993}%
  \BibitemOpen
  \bibfield  {author} {\bibinfo {author} {\bibfnamefont {E.}~\bibnamefont {Hanamura}},\ }\bibfield  {title} {\bibinfo {title} {{Giant two-photon absorption due to excitonic molecule}},\ }\href {https://doi.org/10.1016/0038-1098(93)90297-Z} {\bibfield  {journal} {\bibinfo  {journal} {Solid State Commun.}\ }\textbf {\bibinfo {volume} {88}},\ \bibinfo {pages} {1073} (\bibinfo {year} {1993})}\BibitemShut {NoStop}%
\bibitem [{\citenamefont {Ficek}\ and\ \citenamefont {Tana{\'{s}}}(2002)}]{Ficek2002}%
  \BibitemOpen
  \bibfield  {author} {\bibinfo {author} {\bibfnamefont {Z.}~\bibnamefont {Ficek}}\ and\ \bibinfo {author} {\bibfnamefont {R.}~\bibnamefont {Tana{\'{s}}}},\ }\bibfield  {title} {\bibinfo {title} {{Entangled states and collective nonclassical effects in two-atom systems}},\ }\href {https://doi.org/10.1016/S0370-1573(02)00368-X} {\bibfield  {journal} {\bibinfo  {journal} {Phys. Rep.}\ }\textbf {\bibinfo {volume} {372}},\ \bibinfo {pages} {369} (\bibinfo {year} {2002})}\BibitemShut {NoStop}%
\bibitem [{\citenamefont {Li}\ \emph {et~al.}(2004)\citenamefont {Li}, \citenamefont {Allaart},\ and\ \citenamefont {Lenstra}}]{Li2004a}%
  \BibitemOpen
  \bibfield  {author} {\bibinfo {author} {\bibfnamefont {G.-x.}\ \bibnamefont {Li}}, \bibinfo {author} {\bibfnamefont {K.}~\bibnamefont {Allaart}},\ and\ \bibinfo {author} {\bibfnamefont {D.}~\bibnamefont {Lenstra}},\ }\bibfield  {title} {\bibinfo {title} {{Entanglement between two atoms in an overdamped cavity injected with squeezed vacuum}},\ }\href {https://doi.org/10.1103/PhysRevA.69.055802} {\bibfield  {journal} {\bibinfo  {journal} {Phys. Rev. A}\ }\textbf {\bibinfo {volume} {69}},\ \bibinfo {pages} {055802} (\bibinfo {year} {2004})}\BibitemShut {NoStop}%
\bibitem [{\citenamefont {Prakash}\ and\ \citenamefont {Kumar}(2008)}]{PRAKASH2008}%
  \BibitemOpen
  \bibfield  {author} {\bibinfo {author} {\bibfnamefont {H.}~\bibnamefont {Prakash}}\ and\ \bibinfo {author} {\bibfnamefont {R.}~\bibnamefont {Kumar}},\ }\bibfield  {title} {\bibinfo {title} {Collapses and revivals in two-level atoms in a superposed state interacting with a single mode superposed coherent radiation},\ }\href {https://doi.org/10.1142/S0217979208039800} {\bibfield  {journal} {\bibinfo  {journal} {Int. J. Mod. Phys. B}\ }\textbf {\bibinfo {volume} {22}},\ \bibinfo {pages} {2725} (\bibinfo {year} {2008})}\BibitemShut {NoStop}%
\bibitem [{\citenamefont {Chilingaryan}\ and\ \citenamefont {Rodr{\'{i}}guez-Lara}(2013)}]{Chilingaryan2013}%
  \BibitemOpen
  \bibfield  {author} {\bibinfo {author} {\bibfnamefont {S.~A.}\ \bibnamefont {Chilingaryan}}\ and\ \bibinfo {author} {\bibfnamefont {B.~M.}\ \bibnamefont {Rodr{\'{i}}guez-Lara}},\ }\bibfield  {title} {\bibinfo {title} {{The quantum Rabi model for two qubits}},\ }\href {https://doi.org/10.1088/1751-8113/46/33/335301} {\bibfield  {journal} {\bibinfo  {journal} {J. Phys. A Math. Theor.}\ }\textbf {\bibinfo {volume} {46}},\ \bibinfo {pages} {335301} (\bibinfo {year} {2013})}\BibitemShut {NoStop}%
\bibitem [{\citenamefont {Combescot}\ and\ \citenamefont {Shiau}(2015)}]{Combescot2015}%
  \BibitemOpen
  \bibfield  {author} {\bibinfo {author} {\bibfnamefont {M.}~\bibnamefont {Combescot}}\ and\ \bibinfo {author} {\bibfnamefont {S.-Y.}\ \bibnamefont {Shiau}},\ }\bibfield  {title} {\bibinfo {title} {{Biexcitons}},\ }in\ \href {https://doi.org/10.1093/acprof:oso/9780198753735.003.0014} {\emph {\bibinfo {booktitle} {Excitons and Cooper Pairs}}}\ (\bibinfo  {publisher} {Oxford University Press},\ \bibinfo {year} {2015})\ pp.\ \bibinfo {pages} {340--350}\BibitemShut {NoStop}%
\bibitem [{\citenamefont {Mohamed}\ \emph {et~al.}(2019)\citenamefont {Mohamed}, \citenamefont {Eleuch},\ and\ \citenamefont {Ooi}}]{Mohamed2019a}%
  \BibitemOpen
  \bibfield  {author} {\bibinfo {author} {\bibfnamefont {A.~B.~A.}\ \bibnamefont {Mohamed}}, \bibinfo {author} {\bibfnamefont {H.}~\bibnamefont {Eleuch}},\ and\ \bibinfo {author} {\bibfnamefont {C.~H.~R.}\ \bibnamefont {Ooi}},\ }\bibfield  {title} {\bibinfo {title} {{Non-locality Correlation in Two Driven Qubits Inside an Open Coherent Cavity: Trace Norm Distance and Maximum Bell Function}},\ }\href {https://doi.org/10.1038/s41598-019-55548-2} {\bibfield  {journal} {\bibinfo  {journal} {Sci. Rep.}\ }\textbf {\bibinfo {volume} {9}},\ \bibinfo {pages} {19632} (\bibinfo {year} {2019})}\BibitemShut {NoStop}%
\bibitem [{\citenamefont {Mohamed}\ \emph {et~al.}(2022)\citenamefont {Mohamed}, \citenamefont {ur~Rahman}, \citenamefont {Abdel-Aty}, \citenamefont {Al-Duais},\ and\ \citenamefont {Eleuch}}]{Mohamed2022}%
  \BibitemOpen
  \bibfield  {author} {\bibinfo {author} {\bibfnamefont {A.-B.~A.}\ \bibnamefont {Mohamed}}, \bibinfo {author} {\bibfnamefont {A.}~\bibnamefont {ur~Rahman}}, \bibinfo {author} {\bibfnamefont {A.-H.}\ \bibnamefont {Abdel-Aty}}, \bibinfo {author} {\bibfnamefont {F.~S.}\ \bibnamefont {Al-Duais}},\ and\ \bibinfo {author} {\bibfnamefont {H.}~\bibnamefont {Eleuch}},\ }\bibfield  {title} {\bibinfo {title} {{Quantum memory and coherence dynamics of two qubits interacting with a coherent cavity under intrinsic decoherence}},\ }\href {https://doi.org/10.1007/s11082-022-04192-8} {\bibfield  {journal} {\bibinfo  {journal} {Opt. Quantum Electron.}\ }\textbf {\bibinfo {volume} {54}},\ \bibinfo {pages} {783} (\bibinfo {year} {2022})}\BibitemShut {NoStop}%
\bibitem [{\citenamefont {Eshun}\ \emph {et~al.}(2022)\citenamefont {Eshun}, \citenamefont {Varnavski}, \citenamefont {Villabona-Monsalve}, \citenamefont {Burdick},\ and\ \citenamefont {Goodson}}]{Eshun2022}%
  \BibitemOpen
  \bibfield  {author} {\bibinfo {author} {\bibfnamefont {A.}~\bibnamefont {Eshun}}, \bibinfo {author} {\bibfnamefont {O.}~\bibnamefont {Varnavski}}, \bibinfo {author} {\bibfnamefont {J.~P.}\ \bibnamefont {Villabona-Monsalve}}, \bibinfo {author} {\bibfnamefont {R.~K.}\ \bibnamefont {Burdick}},\ and\ \bibinfo {author} {\bibfnamefont {T.}~\bibnamefont {Goodson}},\ }\bibfield  {title} {\bibinfo {title} {{Entangled Photon Spectroscopy}},\ }\href {https://doi.org/10.1021/acs.accounts.1c00687} {\bibfield  {journal} {\bibinfo  {journal} {Acc. Chem. Res.}\ }\textbf {\bibinfo {volume} {55}},\ \bibinfo {pages} {991} (\bibinfo {year} {2022})}\BibitemShut {NoStop}%
\bibitem [{\citenamefont {Movahedi}\ \emph {et~al.}(2023)\citenamefont {Movahedi}, \citenamefont {Afshar},\ and\ \citenamefont {Jafarpour}}]{Movahedi2023}%
  \BibitemOpen
  \bibfield  {author} {\bibinfo {author} {\bibfnamefont {R.}~\bibnamefont {Movahedi}}, \bibinfo {author} {\bibfnamefont {D.}~\bibnamefont {Afshar}},\ and\ \bibinfo {author} {\bibfnamefont {M.}~\bibnamefont {Jafarpour}},\ }\bibfield  {title} {\bibinfo {title} {{Improvement of the entanglement generation in atomic states using a single-mode field in the Tavis–Cummings model}},\ }\href {https://doi.org/10.1140/epjd/s10053-023-00647-z} {\bibfield  {journal} {\bibinfo  {journal} {Eur. Phys. J. D}\ }\textbf {\bibinfo {volume} {77}},\ \bibinfo {pages} {59} (\bibinfo {year} {2023})}\BibitemShut {NoStop}%
\bibitem [{\citenamefont {Peres}(1996)}]{Peres1996}%
  \BibitemOpen
  \bibfield  {author} {\bibinfo {author} {\bibfnamefont {A.}~\bibnamefont {Peres}},\ }\bibfield  {title} {\bibinfo {title} {{Separability Criterion for Density Matrices}},\ }\href {https://doi.org/10.1103/PhysRevLett.77.1413} {\bibfield  {journal} {\bibinfo  {journal} {Phys. Rev. Lett.}\ }\textbf {\bibinfo {volume} {77}},\ \bibinfo {pages} {1413} (\bibinfo {year} {1996})}\BibitemShut {NoStop}%
\bibitem [{\citenamefont {Horodecki}\ \emph {et~al.}(1996)\citenamefont {Horodecki}, \citenamefont {Horodecki},\ and\ \citenamefont {Horodecki}}]{Horodecki1996}%
  \BibitemOpen
  \bibfield  {author} {\bibinfo {author} {\bibfnamefont {M.}~\bibnamefont {Horodecki}}, \bibinfo {author} {\bibfnamefont {P.}~\bibnamefont {Horodecki}},\ and\ \bibinfo {author} {\bibfnamefont {R.}~\bibnamefont {Horodecki}},\ }\bibfield  {title} {\bibinfo {title} {{Separability of mixed states: necessary and sufficient conditions}},\ }\href {https://doi.org/10.1016/S0375-9601(96)00706-2} {\bibfield  {journal} {\bibinfo  {journal} {Phys. Lett. A}\ }\textbf {\bibinfo {volume} {223}},\ \bibinfo {pages} {1} (\bibinfo {year} {1996})}\BibitemShut {NoStop}%
\bibitem [{\citenamefont {Vidal}\ and\ \citenamefont {Werner}(2002)}]{Vidal2002}%
  \BibitemOpen
  \bibfield  {author} {\bibinfo {author} {\bibfnamefont {G.}~\bibnamefont {Vidal}}\ and\ \bibinfo {author} {\bibfnamefont {R.~F.}\ \bibnamefont {Werner}},\ }\bibfield  {title} {\bibinfo {title} {{Computable measure of entanglement}},\ }\href {https://doi.org/10.1103/PhysRevA.65.032314} {\bibfield  {journal} {\bibinfo  {journal} {Phys. Rev. A}\ }\textbf {\bibinfo {volume} {65}},\ \bibinfo {pages} {032314} (\bibinfo {year} {2002})}\BibitemShut {NoStop}%
\bibitem [{\citenamefont {Horodecki}\ \emph {et~al.}(2009)\citenamefont {Horodecki}, \citenamefont {Horodecki}, \citenamefont {Horodecki},\ and\ \citenamefont {Horodecki}}]{Horodecki2009}%
  \BibitemOpen
  \bibfield  {author} {\bibinfo {author} {\bibfnamefont {R.}~\bibnamefont {Horodecki}}, \bibinfo {author} {\bibfnamefont {P.}~\bibnamefont {Horodecki}}, \bibinfo {author} {\bibfnamefont {M.}~\bibnamefont {Horodecki}},\ and\ \bibinfo {author} {\bibfnamefont {K.}~\bibnamefont {Horodecki}},\ }\bibfield  {title} {\bibinfo {title} {{Quantum entanglement}},\ }\href {https://doi.org/10.1103/RevModPhys.81.865} {\bibfield  {journal} {\bibinfo  {journal} {Rev. Mod. Phys.}\ }\textbf {\bibinfo {volume} {81}},\ \bibinfo {pages} {865} (\bibinfo {year} {2009})}\BibitemShut {NoStop}%
\bibitem [{\citenamefont {Drummond}\ and\ \citenamefont {Gardiner}(1980)}]{Drummond1980a}%
  \BibitemOpen
  \bibfield  {author} {\bibinfo {author} {\bibfnamefont {P.~D.}\ \bibnamefont {Drummond}}\ and\ \bibinfo {author} {\bibfnamefont {C.~W.}\ \bibnamefont {Gardiner}},\ }\bibfield  {title} {\bibinfo {title} {{Generalised P-representations in quantum optics}},\ }\href {https://doi.org/10.1088/0305-4470/13/7/018} {\bibfield  {journal} {\bibinfo  {journal} {J. Phys. A. Math. Gen.}\ }\textbf {\bibinfo {volume} {13}},\ \bibinfo {pages} {2353} (\bibinfo {year} {1980})}\BibitemShut {NoStop}%
\bibitem [{SM()}]{SM}%
  \BibitemOpen
  \href@noop {} {}\bibinfo {note} {See Supplemental Material for discussions on the generalized $P$ representation, photon-mediated electron--electron interactions, and a general cat-state irradiation.}\BibitemShut {Stop}%
\bibitem [{\citenamefont {Spasibko}\ \emph {et~al.}(2017)\citenamefont {Spasibko}, \citenamefont {Kopylov}, \citenamefont {Krutyanskiy}, \citenamefont {Murzina}, \citenamefont {Leuchs},\ and\ \citenamefont {Chekhova}}]{Spasibko2017a}%
  \BibitemOpen
  \bibfield  {author} {\bibinfo {author} {\bibfnamefont {K.~Y.}\ \bibnamefont {Spasibko}}, \bibinfo {author} {\bibfnamefont {D.~A.}\ \bibnamefont {Kopylov}}, \bibinfo {author} {\bibfnamefont {V.~L.}\ \bibnamefont {Krutyanskiy}}, \bibinfo {author} {\bibfnamefont {T.~V.}\ \bibnamefont {Murzina}}, \bibinfo {author} {\bibfnamefont {G.}~\bibnamefont {Leuchs}},\ and\ \bibinfo {author} {\bibfnamefont {M.~V.}\ \bibnamefont {Chekhova}},\ }\bibfield  {title} {\bibinfo {title} {{Multiphoton Effects Enhanced due to Ultrafast Photon-Number Fluctuations}},\ }\href {https://doi.org/10.1103/PhysRevLett.119.223603} {\bibfield  {journal} {\bibinfo  {journal} {Phys. Rev. Lett.}\ }\textbf {\bibinfo {volume} {119}},\ \bibinfo {pages} {223603} (\bibinfo {year} {2017})}\BibitemShut {NoStop}%
\bibitem [{\citenamefont {Manceau}\ \emph {et~al.}(2019)\citenamefont {Manceau}, \citenamefont {Spasibko}, \citenamefont {Leuchs}, \citenamefont {Filip},\ and\ \citenamefont {Chekhova}}]{Manceau2019}%
  \BibitemOpen
  \bibfield  {author} {\bibinfo {author} {\bibfnamefont {M.}~\bibnamefont {Manceau}}, \bibinfo {author} {\bibfnamefont {K.~Y.}\ \bibnamefont {Spasibko}}, \bibinfo {author} {\bibfnamefont {G.}~\bibnamefont {Leuchs}}, \bibinfo {author} {\bibfnamefont {R.}~\bibnamefont {Filip}},\ and\ \bibinfo {author} {\bibfnamefont {M.~V.}\ \bibnamefont {Chekhova}},\ }\bibfield  {title} {\bibinfo {title} {{Indefinite-Mean Pareto Photon Distribution from Amplified Quantum Noise}},\ }\href {https://doi.org/10.1103/PhysRevLett.123.123606} {\bibfield  {journal} {\bibinfo  {journal} {Phys. Rev. Lett.}\ }\textbf {\bibinfo {volume} {123}},\ \bibinfo {pages} {123606} (\bibinfo {year} {2019})}\BibitemShut {NoStop}%
\bibitem [{\citenamefont {Gottesman}\ \emph {et~al.}(2001)\citenamefont {Gottesman}, \citenamefont {Kitaev},\ and\ \citenamefont {Preskill}}]{Gottesman2001}%
  \BibitemOpen
  \bibfield  {author} {\bibinfo {author} {\bibfnamefont {D.}~\bibnamefont {Gottesman}}, \bibinfo {author} {\bibfnamefont {A.}~\bibnamefont {Kitaev}},\ and\ \bibinfo {author} {\bibfnamefont {J.}~\bibnamefont {Preskill}},\ }\bibfield  {title} {\bibinfo {title} {{Encoding a qubit in an oscillator}},\ }\href {https://doi.org/10.1103/PhysRevA.64.012310} {\bibfield  {journal} {\bibinfo  {journal} {Phys. Rev. A}\ }\textbf {\bibinfo {volume} {64}},\ \bibinfo {pages} {012310} (\bibinfo {year} {2001})}\BibitemShut {NoStop}%
\bibitem [{\citenamefont {D'Abbruzzo}\ \emph {et~al.}(2024)\citenamefont {D'Abbruzzo}, \citenamefont {Farina},\ and\ \citenamefont {Giovannetti}}]{DAbbruzzo2024}%
  \BibitemOpen
  \bibfield  {author} {\bibinfo {author} {\bibfnamefont {A.}~\bibnamefont {D'Abbruzzo}}, \bibinfo {author} {\bibfnamefont {D.}~\bibnamefont {Farina}},\ and\ \bibinfo {author} {\bibfnamefont {V.}~\bibnamefont {Giovannetti}},\ }\bibfield  {title} {\bibinfo {title} {{Recovering Complete Positivity of Non-Markovian Quantum Dynamics with Choi-Proximity Regularization}},\ }\href {https://doi.org/10.1103/PhysRevX.14.031010} {\bibfield  {journal} {\bibinfo  {journal} {Phys. Rev. X}\ }\textbf {\bibinfo {volume} {14}},\ \bibinfo {pages} {031010} (\bibinfo {year} {2024})}\BibitemShut {NoStop}%
\end{thebibliography}%
\clearpage
\onecolumngrid
\begin{center}
{\large\bfseries Supplemental Material for\\
``Electron dynamics induced by quantum cat-state light''}\par\vspace{0.8em}
Shohei Imai,$^{1}$ Atsushi Ono,$^{2}$ and Naoto Tsuji$^{1,3}$\par\vspace{0.4em}
{\small
$^{1}$Department of Physics, University of Tokyo, Hongo, Tokyo 113-0033, Japan\\
$^{2}$Department of Physics, Graduate School of Science, Tohoku University, Sendai 980-8578, Japan\\
$^{3}$RIKEN Center for Emergent Matter Science (CEMS), Wako 351-0198, Japan}
\end{center}
\twocolumngrid

\setcounter{section}{0}
\setcounter{equation}{0}
\setcounter{figure}{0}
\renewcommand{\theequation}{S.\arabic{equation}}
\renewcommand{\thefigure}{S.\arabic{figure}}
\providecommand{\theHequation}{}
\providecommand{\theHfigure}{}
\providecommand{\theHsection}{}
\renewcommand{\theHequation}{supp.\arabic{equation}}
\renewcommand{\theHfigure}{supp.\arabic{figure}}
\renewcommand{\theHsection}{supp.\arabic{section}}

\section{Generalized $P$ representation} \label{sup:sec:gene P}
We explain the relationship between the present and previous theories for electron dynamics driven by quantum states of light, the latter proposed by Gorlach \textit{et al.}~\cite{Gorlach2023}.
The latter framework is shown in the Supplementary Information of Ref.~\cite{Gorlach2023}, where they have utilized the generalized $P$ representation~\cite{Drummond1980a} to express the initial nonequilibrium states of light, instead of the Sudarshan--Glauber $P$ representation in our effective theory [Eq.~\eqref{eq:Sudarshan--Glauber P representation}].
In the generalized $P$ representation (or generalized Glauber representation), the photon density matrix is written as
\begin{equation}
\hat{\rho}_{\mathrm{p}}(0) =  \int \mathrm{d}^2\alpha \mathrm{d}^2\beta \, P(\alpha,\beta) \frac{|\alpha\rangle \langle \overline{\beta}|}{\langle \overline{\beta}|\alpha\rangle}, \label{sup:eq:generalized P representation}
\end{equation}
where $P(\alpha,\beta)$ is a positive distribution.

In the absence of radiation from electrons in their theory, one can express the electron density matrix as
\begin{align}
&\hat{\rho}_{\mathrm{e}}^{\mathrm{gene.}\,P}(t) \approx \int \mathrm{d}^2 \alpha \mathrm{d}^2\beta \, P(\alpha,\beta) \hat{\rho}_{\mathrm{e},\alpha,\overline{\beta}}(t), \label{sup:eq:gene. P weighted density matrix} \\
&\mathrm{i}\partial_{t} \hat{\rho}_{\mathrm{e},\alpha,\overline{\beta}}(t) = \hat{\mathcal{H}}_{\mathrm{e}}[\gamma \alpha_0(t)] \hat{\rho}_{\mathrm{e},\alpha,\overline{\beta}}(t) - \hat{\rho}_{\mathrm{e},\alpha,\overline{\beta}}(t) \hat{\mathcal{H}}_{\mathrm{e}}[\gamma \overline{\beta}_0(t)]. \label{sup:eq:EoM}
\end{align}
Here, $\alpha_0(t)=\alpha \exp(-\mathrm{i}\omega t)$ [Eq.~\eqref{eq:free photon field} in the main text] and $\overline{\beta}_0(t) = \overline{\beta} \exp(-\mathrm{i}\omega t)$.
When there are off-diagonal terms of the coherent states in $\hat{\rho}_{\mathrm{p}}(0)$, Eq.~\eqref{sup:eq:EoM} describes how electronic quantum interference arises via the distinct left and right (Hermitian) Hamiltonians.
The difference between the left and right Hamiltonians leads to a time evolution of $\hat{\rho}_{\mathrm{e},\alpha,\overline{\beta}}(t)$ that does not preserve its trace.

There is an ambiguity in Eq.~\eqref{sup:eq:gene. P weighted density matrix} arising from the non-uniqueness of the specific form of $P(\alpha,\beta)$~\cite{Drummond1980a}.
As an example, let us consider the coherent state $|\alpha_0\rangle$.
One of the expressions for $P(\alpha,\beta)$ is given by $P(\alpha,\beta)=\frac{1}{4\pi}\exp(-|\alpha-\overline{\beta}|^2/4) \, Q((\alpha+\overline{\beta})/2)$ with the Husimi function $Q(\alpha)=\langle \alpha|\hat{\rho}_{\mathrm{p}} |\alpha\rangle/\pi$.
From this formula, one obtains 
\begin{equation}
P(\alpha,\beta) = \frac{1}{4\pi^2} \mathrm{e}^{-|\alpha-\alpha_0|^2/2} \, \mathrm{e}^{-|\overline{\beta}-\alpha_0|^2/2}. \label{sup:eq:coherent state:exp}
\end{equation}
Another representation is given by
\begin{equation}
P(\alpha,\beta) = \delta^2(\alpha-\alpha_0)\delta^2(\overline{\beta}-\alpha_0). \label{sup:eq:coherent state:delta}
\end{equation}
When calculating the photon density matrix~\eqref{sup:eq:generalized P representation} and an expectation value of a normal-ordered operator, one obtains the same results for both representations.
To see this, let us first consider an expectation value of $\hat a^{\dagger}{}^m \hat a^n$, which is expressed as an integral form as follows:
\begin{equation}
\mathrm{Tr}_{\mathrm{p}}\left[ \hat{a}^{\dagger}{}^m\, \hat{a}^n\, \hat{\rho}_{\mathrm{p}} \right] = \int \mathrm{d}^2\alpha  \mathrm{d}^2\beta \,\beta^m\alpha^n\,P(\alpha,\beta). \label{sup:eq:expectation value for normal-ordered operator}
\end{equation}
Here, the following formula holds for any analytic function $f(\alpha)$ (satisfying the Cauchy--Riemann equation $\partial f/\partial \overline{\alpha} = 0$):
\begin{equation}
f(\alpha') = \int \frac{\mathrm{d}^2\alpha}{2\pi} f(\alpha) \mathrm{e}^{-|\alpha-\alpha'|^2/2}. \label{sup:eq:analytic function gaussian integral}
\end{equation}
Applying Eq.~\eqref{sup:eq:analytic function gaussian integral} to the analytic function $\beta^m\alpha^n$ in Eq.~\eqref{sup:eq:expectation value for normal-ordered operator}, we can derive the expectation value $\langle \hat{a}^{\dagger}{}^m \hat{a}^n \rangle = \overline{\alpha_0}^m \alpha_0^n$ for the coherent state $|\alpha_0 \rangle$, regardless of whether $P(\alpha,\beta)$ is expressed as in Eq.~\eqref{sup:eq:coherent state:exp} or~\eqref{sup:eq:coherent state:delta}.
Furthermore, since the kernel $|\alpha\rangle \langle \overline{\beta}|/\langle \overline{\beta}|\alpha\rangle$ in Eq.~\eqref{sup:eq:generalized P representation} is an analytic function of both $\alpha$ and $\beta$, the photon density matrix $\hat{\rho}_{\mathrm{p}}(0)$ remains the same no matter what kind of the form of $P(\alpha, \beta)$ is used.
Therefore, the non-uniqueness of the specific forms of $P(\alpha,\beta)$ does not affect the photon density matrix and the expectation values of any normal-ordered operators.

However, the operator $\hat{\rho}_{\mathrm{e},\alpha,\overline{\beta}}(t)$ in Eq.~\eqref{sup:eq:gene. P weighted density matrix} is not an analytic function of $\alpha$ and $\beta$ in general, so that a deviation may occur depending on the forms of $P(\alpha,\beta)$.
This deviation comes from the fact that the time-evolution operators $\hat{\mathrm{T}}_{\pm}\exp[\mp \mathrm{i} \int \mathrm{d}\tau \hat{\mathcal{H}}_{\mathrm{e}}[\gamma\hat{a}]]$ are not normal ordered with respect to $\hat{a}$.
Generally, for a non-normal-ordered operator, there is no straightforward way to express its expectation value in an integral form, such as Eq.~\eqref{sup:eq:expectation value for normal-ordered operator}, and we have to expand the operator to be normal-ordered ones.
However, in the case of time-ordered operators, such as the time-evolution operators, the path integral representation is valid.
Our theory starts from this established treatment, and we adopt the Born approximation for the path integral.
For the form of $P(\alpha,\beta)$, Eq.~\eqref{sup:eq:coherent state:delta} might be more suitable than Eq.~\eqref{sup:eq:coherent state:exp}, since the former reproduces the conventional well-established formulation that considers the time-evolution of a pure state under laser irradiation.

In the cat-state light irradiation represented by Eq.~\eqref{eq:cat state}, the generalized $P$ function is given by
\begin{align}
P(\alpha,\beta) =\frac{1}{\mathcal{N}_{\alpha_0}} \Bigl\{
  &\delta^2(\alpha - \alpha_0)\delta^2(\overline{\beta} - \alpha_0) \nonumber \\
+ &\delta^2(\alpha + \alpha_0)\delta^2(\overline{\beta} + \alpha_0) \nonumber \\
+ &\langle -\alpha_0|\alpha_0 \rangle \delta^2(\alpha - \alpha_0)\delta^2(\overline{\beta} + \alpha_0) \nonumber \\
+ &\langle -\alpha_0|\alpha_0 \rangle \delta^2(\alpha + \alpha_0)\delta^2(\overline{\beta} - \alpha_0) 
\Bigr\}. \label{eq:cat state:gene P}
\end{align}
Through the perturbative analysis [Eq.~\eqref{eq:perturb:one-body state}], we obtain the interferential components of the two-body electron density matrix as
\begin{equation}
\hat{\rho}_{2,\mathrm{I}}^{\mathrm{gene.}\, P, \mathrm{itf}} = \frac{\langle -\alpha_0|\alpha_0\rangle}{\mathcal{N}_{\alpha_0}} \left( |\psi_+^{\mathrm{cls}}\rangle\langle\psi_-^{\mathrm{cls}}| + |\psi_-^{\mathrm{cls}}\rangle\langle\psi_+^{\mathrm{cls}}| \right). \label{sup:eq:two-body density matrix:itf}
\end{equation}
Here, the classically driven contribution is the same as in Eq.~\eqref{eq:two-body density matrix:cls:state vector} in the main text.
We can clearly see the deviation between Eq.~\eqref{sup:eq:two-body density matrix:itf} and Eq.~\eqref{eq:two-body density matrix:itf:state vectors}, and our results [Eq.~\eqref{eq:two-body density matrix:itf:state vectors}] correctly reproduce the full-system simulations, as shown in the inset of Fig.~\ref{fig:cat}(b).
Using a different representation of the generalized $P$ function given by $P(\alpha,\beta)= [\delta^2(\alpha{-}\alpha_0) +\delta^2(\alpha{+}\alpha_0) +\frac{1}{4\pi^2}\langle-\alpha_0|\alpha_0\rangle (\mathrm{e}^{-|\alpha{+}\alpha_0|^2/2}\,
\mathrm{e}^{\overline{\alpha_0}(\alpha{+}\alpha_0)}\,
\mathrm{e}^{-|\beta{-}\overline{\alpha_0}|^2/2}\,
\mathrm{e}^{-(\beta{-}\overline{\alpha_0})\alpha_0}
+\mathrm{e}^{-|\alpha{-}\alpha_0|^2/2}\,
\mathrm{e}^{-\overline{\alpha_0}(\alpha{-}\alpha_0)}\,
\mathrm{e}^{-|\beta{+}\overline{\alpha_0}|^2/2}\,
\mathrm{e}^{(\beta{+}\overline{\alpha_0})\alpha_0}) ]/\mathcal{N}_{\alpha_0}$, in which the diagonal components of the coherent states are expressed as delta functions and the off-diagonal components as Husimi functions, we found that the deviation becomes $3$--$4$ times larger than that observed using Eq.~\eqref{eq:cat state:gene P} (not shown).

When a time-evolving electronic state is expressed as an analytic function of $\alpha$ and $\beta$, we can see the consistency between our effective theory [Eqs.~\eqref{eq:P weighted density matrix} and~\eqref{eq:EoM}] and the generalized $P$-representation theory [Eqs.~\eqref{sup:eq:gene. P weighted density matrix} and~\eqref{sup:eq:EoM}].
We will use the following relationship between the Sudarshan--Glauber $P$ function, denoted as $P(\alpha)$, and the generalized $P$ function, $P(\alpha,\beta)$~\cite{Drummond1980a}:
\begin{equation}
P(\alpha, \beta) = \int \frac{\mathrm{d}^2\alpha'}{4\pi^2} P(\alpha') \mathrm{e}^{-|\alpha-\alpha'|^2/2}\, \mathrm{e}^{-|\overline{\beta}-\alpha'|^2/2}. \label{sup:eq:from P to P(a,b)}
\end{equation}
Let the initial electronic state $\hat{\rho}_{\mathrm{e}}(0)$ be a pure state $|\psi_{\mathrm{e}}(0)\rangle$, and the time-evolving electronic state $|\psi_{\mathrm{e},\alpha}\rangle \equiv \hat{\mathrm{T}}_+ \mathrm{e}^{-\mathrm{i}\int_0^t\mathrm{d}\tau \hat{\mathcal{H}}_{\mathrm{e}}[\gamma \alpha_0(\tau)]} |\psi_{\mathrm{e}}(0)\rangle$ be an analytic function of $\alpha$ via $\alpha_0(\tau)=\alpha \exp(-\mathrm{i}\omega \tau)$.
Substituting Eq.~\eqref{sup:eq:from P to P(a,b)} into Eq.~\eqref{sup:eq:gene. P weighted density matrix}, we obtain the electron density matrix as
\begin{widetext}
\begin{align}
\hat{\rho}_{\mathrm{e}}(t) \approx&  \int \mathrm{d}^2\alpha' P(\alpha')
\left[ \int\frac{\mathrm{d}^2\alpha}{2\pi} \mathrm{e}^{-|\alpha-\alpha'|^2/2}\, \hat{\mathrm{T}}_+ \mathrm{e}^{-\mathrm{i}\int_0^t\mathrm{d}\tau \hat{\mathcal{H}}_{\mathrm{e}}[\gamma \alpha_0(\tau)]} |\psi_{\mathrm{e}}(0)\rangle \right] \, 
\left[\int\frac{\mathrm{d}^2\beta}{2\pi} \mathrm{e}^{-|\overline{\beta}-\alpha'|^2/2}\, \langle \psi_{\mathrm{e}}(0)|\hat{\mathrm{T}}_- \mathrm{e}^{\mathrm{i}\int_0^t\mathrm{d}\tau \hat{\mathcal{H}}_{\mathrm{e}}[\gamma \overline{\beta}_0(\tau)]} \right]
\label{sup:eq:from gene P to SG P in electron density matrix} \\
= & \int \mathrm{d}^2\alpha' P(\alpha') |\psi_{\mathrm{e},\alpha'}\rangle \langle \psi_{\mathrm{e},\alpha'}|. \label{sup:eq:SG P for pure state}
\end{align}
\end{widetext}
Equation~\eqref{sup:eq:SG P for pure state} is equivalent to that of our effective theory when the initial state is a pure state [Eqs.~\eqref{eq:P weighted density matrix} and~\eqref{eq:EoM}].
However, it is not generally true that the time-evolving electronic state $|\psi_{\mathrm{e},\alpha}\rangle = \hat{\mathrm{T}}_+ \mathrm{e}^{-\mathrm{i}\int_0^t\mathrm{d}\tau \hat{\mathcal{H}}_{\mathrm{e}}[\gamma \alpha_0(\tau)]} |\psi_{\mathrm{e}}(0)\rangle$ becomes an analytic function of $\alpha$.
Therefore, we have to numerically check which theory can correctly reproduce the full system simulations, which is discussed in the main text and is shown in the inset of Fig.~\ref{fig:cat}(b).

Finally, we comment on the compatibility of the Born approximation.
In the Sudarshan--Glauber $P$ representation, we can consider the closed trajectories, namely $\alpha_{+}(\tau) = \alpha_{-}(\tau)$ at the boundaries $\tau=0$ and $t$, in the path integral in Eq.~\eqref{eq:path integral}.
This enables us to apply the Born approximation in a straightforward way.
In contrast, in the generalized $P$ representation, we have to evaluate a path integral for open trajectories, namely $\alpha_+(0)=\alpha$ and $\alpha_-(0)=\overline{\beta}$, which prevents us from directly applying the Born approximation.

\section{Photon-mediated electron--electron interactions}
 \label{sup:sec:1st Born}
We discuss the backaction effects of electrons on photons by considering the next order of the Born approximation.
The equation of motion for the disturbed photons is given by Eq.~\eqref{eq:1st-order photon EoM}, whose formal solution is written as
\begin{equation}
\alpha_1(\tau) =  \alpha_0(\tau) +\mathrm{i} \gamma \int_{0}^{\tau}\mathrm{d}\tau' \hat{\mathcal{J}}^-[\gamma \alpha_0(\tau')]\, \mathrm{e}^{-\mathrm{i}\omega (\tau-\tau')}, \label{sup:eq:1st-order photon solution}
\end{equation}
which is referred to as the first Born approximation.

Using this modulated photon trajectory $\alpha_1(\tau)$, we can express the time-evolution operator $\hat{U}_{1\mathrm{BA}}(t,0)$ of the electron system as
\begin{widetext}
\begin{align}
\hat{U}_{1\mathrm{BA}}(t,0) &= \hat{\mathrm{T}} \exp \left\{ -\mathrm{i} \int_0^t \mathrm{d}\tau \hat{\mathcal{H}}_{\mathrm{e}}[\gamma \alpha_1(\tau)]
\right\} \label{sup:eq:1st Born time evolution operator}\\
&= \hat{\mathrm{T}} \exp \left( -\mathrm{i} \int_0^t \mathrm{d}\tau 
\hat{\mathcal{H}}_{\mathrm{e}}[\gamma \alpha_0(\tau)]  
-\mathrm{i} \int_0^t \mathrm{d}\tau \int_{0}^{t} \mathrm{d}\tau'
\Bigl\{ -\mathrm{i} \gamma^2 \hat{\mathcal{J}}^-[\gamma \alpha_0(\tau)] 
\hat{\mathcal{J}}^+[\gamma \alpha_0(\tau')]\, \mathrm{e}^{-\mathrm{i}\omega(\tau - \tau')}
+{\mathrm{H.c.}} \Bigr\}
\right). \label{sup:eq:1st Born time evolution operator:current-current}
\end{align}
\end{widetext}
The second term in the parentheses in Eq.~\eqref{sup:eq:1st Born time evolution operator:current-current} represents the photon-mediated electron--electron interactions involving current fluctuations.

We apply the mean-field approximation to the current--current interaction term.
The laser irradiation induces a finite expectation value of the electric current $\langle \hat{\mathcal{J}} \rangle$ proportional to the external vector potential $A_0(t)=\gamma[\alpha_0(t)+\overline{\alpha_0(t)}]$ in the linear response.
This is regarded as a vector potential deviation $\updelta A =A_1 - A_0 \sim \gamma^3 \alpha_0$, where $A_1(t) = \gamma [\alpha_1(t)+ \overline{\alpha_1(t)} ]$.
This $\gamma$ dependence leads to the trace distance scaling via Eq.~\eqref{sup:eq:1st Born time evolution operator:current-current}, shown as the red line in Fig.~\ref{fig:gamma}.

Next, let us consider the cat-state light irradiation.
The expectation value of the current $\langle \hat{\mathcal{J}} \rangle$ vanishes, but the current fluctuation $\langle \hat{\mathcal{J}} \hat{\mathcal{J}} \rangle$ becomes finite.
From Eqs.~\eqref{eq:two-body density matrix:cls:matrix form} and~\eqref{eq:two-body density matrix:itf:matrix form}, 
the expectation value $\langle \hat{\mathcal{J}} \hat{\mathcal{J}} \rangle$ is proportional to $\gamma^2$, making the interaction energy~\eqref{sup:eq:1st Born time evolution operator:current-current} proportional to $\gamma^4$, which is consistent with the blue line in Fig.~\ref{fig:gamma}.

\section{General cat-state light irradiation}
 \label{sup:sec:general_cat}
\begin{figure}[tb]
\centering
\includegraphics[width=\columnwidth]{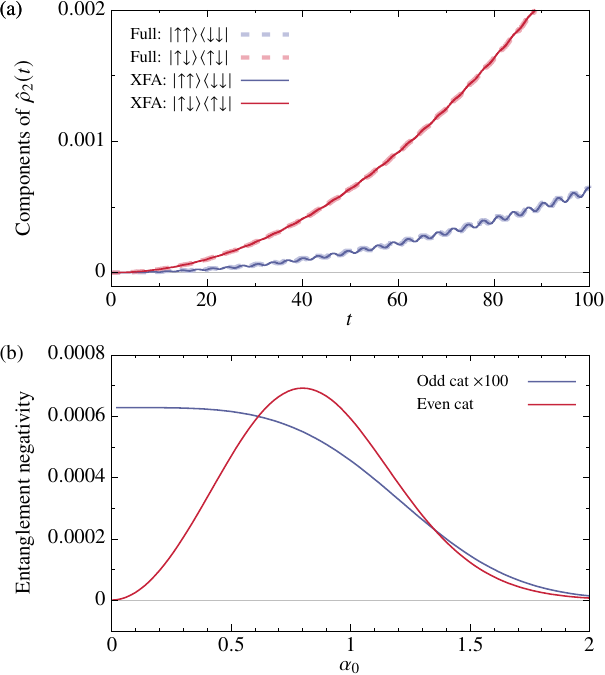}
\caption{Electron dynamics induced by odd cat-state light.
(a)~Real-time evolution of the two-body electron density matrix $\hat{\rho}_2(t)$.
(b)~Entanglement negativity between electrons.
Even cat result (red curve) is the same as Fig.~\ref{fig:cat}(b) in the manuscript.
Parameters are the same as those used in the main text.
}
\label{sup:fig:odd_cat}
\end{figure}
We examine electron dynamics induced by a general class of cat states, in which a relative phase $\theta$ is introduced between the two coherent states:
\begin{equation}
|\mathrm{cat}_\theta\rangle =\frac{1}{\sqrt{\mathcal{N}_{\alpha_0,\theta}}} \left(|\alpha_0\rangle +\mathrm{e}^{\mathrm{i}\theta}|{-\alpha}_0 \rangle \right), \label{sup:eq:general cat state}
\end{equation}
where $\mathcal{N}_{\alpha_0,\theta}$ is a normalization constant.
The induced electron dynamics can be described using our effective theory [Eqs.~\eqref{eq:P weighted density matrix} and~\eqref{eq:EoM}] as in the even cat case [Eq.~\eqref{eq:cat state}, $\theta = 0$], by employing the corresponding Sudarshan--Glauber $P$ function that includes the relative phase:
\begin{align}
&P_{\mathrm{cat},\theta}(\alpha) = \frac{1}{\mathcal{N}_{\alpha_0,\theta}} \Bigl\{ \delta^2(\alpha - \alpha_0) + \delta^2(\alpha + \alpha_0) \nonumber \\
&+ \langle -\alpha_0|\alpha_0 \rangle \left[ \mathrm{e}^{-\mathrm{i}\theta}\,\tilde{\delta}(\alpha{-}\alpha_0)\tilde{\delta}(\overline{\alpha}{+}\overline{\alpha_0}) + \mathrm{e}^{\mathrm{i}\theta}\,\tilde{\delta}(\alpha{+}\alpha_0)\tilde{\delta}(\overline{\alpha}{-}\overline{\alpha_0}) \right] \Bigr\}. \label{sup:eq:cat state:P}
\end{align}

As an important case, we now focus on the odd cat state, corresponding to $\theta = \pi$.
Figure~\ref{sup:fig:odd_cat}(a) shows the time evolution of the reduced two-electron density matrix under odd cat state irradiation, using the same setup as in the main text.
We confirm that the results obtained by the exact full-system simulation (dashed curves) and our effective theory (XFA, solid curves) are in good agreement.
In contrast to the even cat case, we observe that the population growth in $\hat{\rho}_2(t)$ is dominated not by the biexciton component ($|{\uparrow\uparrow}\rangle \langle{\downarrow\downarrow}|$, blue curves), but rather by the single-exciton component ($|{\uparrow\downarrow}\rangle \langle{\uparrow\downarrow}|$, red curves).
Meanwhile, since the expectation value of the electric field vanishes ($\langle \mathrm{cat}_\pi| \hat{E} | \mathrm{cat}_\pi \rangle = 0$), the expectation values of the electric polarization and current also vanish: $\langle \hat{S}_{1,2}^{\pm} \rangle = 0$.
Figure~\ref{sup:fig:odd_cat}(b) shows the entanglement negativity $\mathcal{N}$, which remains finite even for $\alpha_0 \to 0$.
Quantitatively, the value is roughly two orders of magnitude smaller than in the even cat state case.

To clarify the physical interpretation of this dynamics, we examine a perturbative analysis similar to that in the main text.
In the rotating-wave and weak-excitation approximations, the reduced electron state under the odd-cat-state irradiation is given by $\hat{\rho}_{2,\mathrm{I}} = \hat{\rho}_{2,\mathrm{I}}^{\mathrm{cls}} - \hat{\rho}_{2,\mathrm{I}}^{\mathrm{itf}} \approx |{\downarrow\downarrow}\rangle\langle{\downarrow\downarrow}| + |\mu\omega\gamma t|^2/2 \cdot (|{\uparrow\downarrow}\rangle + |{\downarrow\uparrow}\rangle)(\langle{\uparrow\downarrow}| + \langle{\downarrow\uparrow}|)$.
This is a mixed state containing an entangled single-exciton component ($|{\uparrow\downarrow}\rangle + |{\downarrow\uparrow}\rangle$), and its amplitude does not vanish in the limit $\alpha_0 \to 0$.
This accounts for the persistent entanglement observed in Fig.~\ref{sup:fig:odd_cat}(b).
These behaviors are different from those observed in the even-cat-state case, where the electrons are in a pure biexciton state  $|{\downarrow\downarrow}\rangle + |\mu\omega\gamma\alpha_0 t|^2 |{\uparrow\uparrow}\rangle$.

Furthermore, when the relative phase $\theta$ is neither $0$ nor $\pi$, the expectation value of the electric field becomes finite as $\langle \mathrm{cat}_\theta | \hat{E} | \mathrm{cat}_\theta \rangle = 2 \gamma \omega (\alpha_0 + \overline{\alpha}_0 ) \sin{\theta} \cdot \langle \alpha_0 | {-}\alpha_0 \rangle / \mathcal{N}_{\alpha_0,\theta}$, resulting in finite electric polarization and current generation according to classical electrodynamics.
However, the amplitude is expected to be very small, as it is scaled by the overlap $\langle \alpha_0 | {-}\alpha_0 \rangle$ between the two coherent states.


\end{document}